\documentclass{article}

\usepackage{arxiv}
\usepackage{eurosym}
\usepackage[utf8]{inputenc} % allow utf-8 input
\usepackage[T1]{fontenc}    % use 8-bit T1 fonts
\usepackage{hyperref}       % hyperlinks
\usepackage{url}            % simple URL typesetting
\usepackage{booktabs}       % professional-quality tables
\usepackage{amsfonts}       % blackboard math symbols
\usepackage{nicefrac}       % compact symbols for 1/2, etc.
\usepackage{microtype}      % microtypography
\usepackage{lipsum}
\usepackage{graphicx}
\usepackage{color}
\usepackage{amssymb}
\usepackage[ruled,vlined]{algorithm2e}
\usepackage{subcaption}
\usepackage{amsmath}
\usepackage[autostyle]{csquotes}

\title{Smart non-intrusive appliance identification using a novel local power histogramming descriptor with an improved k-nearest neighbors classifier}

\author{
  Yassine Himeur\thanks{Sustainable Cities and Society 67, 102764, 2021} , Abdullah Alsalemi, Faycal Bensaali\\
  Department of Electrical Engineering\\
  Qatar University\\
  Doha, Qatar \\
  \texttt{yassine.himeur@qu.edu.qa;a.alsalemi@qu.edu.qa;f.bensaali@qu.edu.qa} \\
   \And
 Abbes Amira \\
  Institute of Artificial Intelligence\\
  De Montfort University\\
  Leicester, United Kingdom \\
  \texttt{abbes.amira@dmu.ac.uk} \\
}

\begin{document}
\maketitle

\begin{abstract}
Non-intrusive load monitoring (NILM) is a key cost-effective technology for monitoring power consumption and contributing to several challenges encountered when transiting to an efficient, sustainable, and competitive energy efficiency environment. This paper proposes a smart NILM system based on a novel local power histogramming (LPH) descriptor, in which appliance power signals are transformed into 2D space and short histograms are extracted to represent each device. Specifically, short local histograms are drawn to represent individual appliance consumption signatures and robustly extract appliance-level data from the aggregated power signal. Furthermore, an improved k-nearest neighbors (IKNN) algorithm is presented to reduce the learning computation time and improve the classification performance. This results in highly improving the discrimination ability between appliances belonging to distinct categories. A deep evaluation of the proposed LPH-IKNN based solution is investigated under different data sets, in which the proposed scheme leads to promising performance. An accuracy of up to 99.65\% and 98.51\% has been achieved on GREEND and UK-DALE data sets, respectively. While an accuracy of more than 96\% has been attained on both WHITED and PLAID data sets. This proves the validity of using 2D descriptors to accurately identify appliances and create new perspectives for the NILM problem.
\end{abstract}

% keywords can be removed

\section{Introduction} \label{sec1}
Buildings are responsible on more than 32 percent of the overall energy consumed worldwide, and this percentage is expected to be doubled by 2050 as a result of the well-being improvement and wide use of electrical appliances and central heating/cooling systems \cite{Elattar2020}. Specifically, this is due to population growth, house comfort enhancement and improvement of wealth and lifestyle. To that end, reducing wasted energy and promoting energy-saving in buildings have been nowadays emerged as a hot research topic. One of the cost-effective solutions is via encouraging energy-efficiency behaviors among building end-users based on analyzing energy consumption footprints of individual appliances. Therefore, tailored recommendations can be generated to help end-users improve their behavior \cite{sardianos2020emergence}.

In this regard, load monitoring of appliances can not only provide the end-users with their fine-grained consumption footprints, but it promptly contributes in promoting sustainability and energy efficiency behaviors \cite{REHABC2020}. Moreover, it can significantly contribute in elaborating and developing reliable smart-grid demand management systems. On the other hand, load consumption monitoring principally encompasses two wide groups, namely intrusive load monitoring (ILM) and non-intrusive load monitoring (NILM), respectively. ILM necessitates to install smart-meters at the front-end of each electrical appliance aiming at collecting real-time energy consumption patterns. Even though this strategy presents high performance in accurately gathering appliance-specific data, it requires a heavy lifting with high-cost installation, where a large number of sun-meters are installed. In addition an intrusive transformation of the available power grid is essential \cite{alsalemi2020micro}. On the contrary, no additional sub-meter required when the NILM strategy (named energy disaggregation as well) is adopted to infer device-specific consumption footprints since the latter are immediately extracted from the main load using feature extraction and learning models \cite{HIMEUR2020114877}.

In this context, the NILM issue has been investigated for many years and extensive efforts are still paid to this problematic because of its principal contributions to improve energy consumption behavior of end-users \cite{PEREIRA2020102399,LIU2020101918}. Specifically, it can help in achieving a better comprehending of consumers' consumption behavior through supplying them with specific appliance data. Therefore, put differently, the NILM task indirectly aims at (i) promoting the energy efficiency behavior of individuals, (ii) reducing energy bills and diminishing reliance on fossil fuel, and (iii) reducing carbon emissions and improving environmental conditions \cite{alsalemi2020achieving}.

Two crucial stages in NILM are the feature extraction and inference and learning procedures. The feature extraction step aims at deriving pertinent characteristics of energy consumption signals to help representing appliances from the same category with similar signatures while differencing between power signals from different classes \cite{Welikala8039522}. On the other hand, the inference and learning step is essentially reserved to train classifiers in order to identify appliances and extract appliance-level power footprints \cite{He8669739}. It can be achieved either by using conventional classification models, such as artificial neural networks (ANN), support vector machine (SVM), k-nearest neighbors (KNN), etc. or novel classifiers, including deep neural networks (DNNs). Consequently, the identification of electrical devices simultaneously operating through an interval of time in a household is the central part of the NILM architecture. Its performance is highly dependent on the deployed feature extraction and inference model. To that end, the development of robust schemes belonging to these two modules attracts a considerable interest in recent years \cite{Park2019,Ma8118142}.

In this paper, recent NILM systems are first reviewed based on the principal components contributing into the implementation of such architectures including feature extraction and learning models. In this respect, techniques pertaining to three main feature extraction categories are described among them graph signal processing (GSP), sparse coding features and binary encoding schemes. Following, a discussion of their limitations and drawbacks is also presented after conducting a deep comparison of their performances and properties. Moving forward, a non-intrusive appliance identification architecture is proposed, which is mainly based on a novel local power histogramming (LPH) descriptor. The latter relies on (i) representing power signals in a 2D space, (ii) performing a binary power encoding in small regions using square patches of $3 \times 3$ samples and (iii) returning back to the initial 1D space through extracting histograms of 2D representations. Following, an improved k-nearest neighbors (IKNN) is introduced to effectively identify appliance-level fingerprints and reduce the computation cost. This has resulted in very short appliance signatures of 256 samples, in which each power signal is represented with a unique histogram, and thus leads to a better appliance identification performance at a low computational complexity. Moreover, it is worth-noting that to the best of the authors' knowledge, this paper is the first work that discusses the applicability of 2D local descriptors for identifying electrical appliances using their power consumption signals. Overall, The main contributions of this paper can be summarized as follows:

\begin{itemize}
\item We present a comprehensive overview of recent trends in event-based NILM systems along with describing the their drawbacks and limitations.
\item We propose a novel NILM framework based on an original 2D descriptor, namely LPH, which can be considered as an interesting research direction to develop robust and reliable NILM solutions. Explicitly, after converting appliance power signals into 2D space, the appliance identification becomes a content-based image retrieval (CBIR) problem and a powerful short description is extracted to represent each electrical device. According, LPH operates also as a dimensionality reduction, where each resulted appliance signature has only 256 samples.
\item We design a powerful IKNN model that efficiently aids in recognizing appliances from the extracted LPH fingerprints and reducing significantly the computational cost.
\item We evaluate the performance of the proposed LPH-IKNN based NILM system on four different data sets with distinct sampling frequency rates and in comparison with various recent NILM systems and other 2D descriptors.  
\end{itemize}

The remainder of this paper is structured as follows. An overview of NILM systems is introduced in Section \ref{sec2} along with a discussion of their drawbacks and limitations. In Section \ref{sec3}, the main steps of the proposed NILM system based on the LPH descriptor and IKNN are described in details. The performance results of the exhaustive empirical evaluation conducted in this framework are presented and thoroughly discussed in Section \ref{sec4}, in which different comparisons are conducted with state-of-the-art works. Finally, Section \ref{sec5} concludes the paper, discusses the important findings and highlights the future works.

\section{Related work} \label{sec2}
\subsection{Overview of NILM techniques}
NILM frameworks can be categorized into two major groups. The first one called non-event-based approaches, which focus on using algorithms without depending on the training/learning procedures (using data from a particular building). They can segregate the main power signal collected from the overall circuit into various appliance-level fingerprints. An explicit example of this kind of techniques that have been typically studied is related the deployment of statistical analysis, including hidden Markov models (HMM) \cite{Makonin7317784}, higher-order statistics (HOS) \cite{Guedes2015} or probabilistic models \cite{Ji8684887}. The second group deals with methods allowing the identification of state changes occurred in power consumption signals using different types event detectors, classifiers, and further implementing appropriate techniques to calculate an individual load usage fingerprint for each electrical device. In this section, we focus on describing recent NILM systems pertaining to the second category because the proposed framework is an even-based NILM framework.

Explicitly, this category of NILM systems deploys two principal components. The first one is a feature descriptor to extract pertinent characteristics of electrical appliances, while the second is a learning algorithm that can help in detecting and classifying each device based on its features. Conventional NILM methods have been basically concentrated on extracting features related to steady-states and transient-states, in addition to the adoption of conventional machine learning (ML) classifiers. On the other side, novel strategies are introduced in recent years to deal with the NILM issue based on the use of new signal analysis procedures and innovative learning models. This class of NILM frameworks is defined as non-conventional, they are classified into four principal sub-categories as follows: 
\vskip2mm
\noindent \textbf{Graph signal processing (GSP):} A trending research field aiming at describing stochastic characteristics of power signals based on graph theory. 
In \cite{He7539273}, a graph-based method for identifying individual appliances has been introduced after detecting appliance events. This results in a better detection of appliance-level fingerprints and further a reduction of time computation compared to conventional graph-based techniques. 
In \cite{Li8437176}, various multi-label graphs have been developed to detect individual devices based on a semi-supervised procedure. In \cite{Zhao2018Access}, NILM performance have been enhanced via the use of a generic GSP-based technique, which is build upon the application  of graph-based filters. This results in a better detection of on/off appliance states, via the mitigation of electric noise produced by appliances.

\vskip2mm

\noindent \textbf{Sparse coding features:} In this category, the NILM framework is treated as a blind source separation problem and recent sparse coding schemes are then applied to split an aggregated power consumption signal into specific appliance based profiles \cite{Kolter2010EDV}. In \cite{Singh2019SG}, a co-sparse analysis dictionary learning is proposed to segregate the total energy consumption into a device-level data and significantly shorten the training process. In \cite{Singh7847445}, a deep learning architecture is used for designing a multi-layer dictionary of each appliance rather than constructing one-level codebook. Obtained multi-layer codebooks are then deployed as features for the source-separation algorithm in order to break down the aggregated energy signal. In \cite{Rahimpour7835299}, an improved non-negative matrix factorization is used to pick up perceptibly valuable appliance-level signatures from the aggregated mixture.

\vskip2mm

\noindent \textbf{Binary descriptions:} Most recently, binary descriptors have been investigated for the classification and fault detection of 1D signals such as electroencephalogram (EEG), electrocardiogram (ECG), and myoelectric signals \cite{HAMMAD2019180}. For power consumption signals, this concept is novel. The only few works that can be found in the literature are mainly focusing on representing the power signal in a novel space and directly being used to train the convolutional neural network (CNN). In \cite{Du7130652}, power fingerprints are derived by estimating the similarity of voltage-current (V-I) shapes, encoding it using a binary dictionary and then extracting image graphical footprints that are directly fed to a self-organizing map (SOM) classifier, which is based on neural networks. In \cite{Gao7418189}, V-I binary representation is employed through converting the normalized V-I magnitude into binary matrices using a thresholding process before being fed to a CNN. More specifically, this approach relies on binary coding of the V-I edges plotted in the new representation. These data are then fed into an ML classifier in order to identify each appliance class. In \cite{Liu8580416}, a color encoder is proposed to draw V-I signatures that can also be translated to visual plots. These footprints are then fed to a deep learning classifier to identify each electrical appliance. In \cite{DEBAETS2019645}, a siamese-neural network is employed aims at mapping the V-I trajectories into a novel characteristic representation plan. 

\vskip2mm

\noindent \textbf{Time-frequency analysis:} Time-frequency analysis is an imperative research topic, in which much attention has been devoted to it in the past and even nowadays. It is applied in several applications among them energy efficiency \cite{JUNKER2018175}, NILM or energy disaggregation \cite{Himeur2020icict} and power consumption anomaly detection \cite{WANG2020114145}. In \cite{himeur2020effective,Himeur2020iscas}, a novel NILM descriptor is proposed based on the fusion of different time-domain descriptors. In \cite{HIMEUR2020114877}, a novel time-scale analysis is adopted based on the use of multi-scale wavelet packet trees (MSWPT) and a cepstrum-based event detection scheme to glean appliance-level power consumption patterns from the aggregated load.

\subsection{Classification}
\subsubsection{Improved k-nearest neighbors}  \label{sec221}
In the building energy sector, KNN has been widely deployed in the literature for different purposes, such as energy disaggregation \cite{shi2019nonintrusive} and anomaly detection \cite{Himeur2020IJIS-AD,mulongo2020anomaly,himeur2020novel} although it has some issues, e.g. the sensitivity of the neighborhood size $k$ could significantly degrade its performance \cite{mehta2018new,abu2019effects}. To that end, an improved version is proposed in \cite{gou2019generalized} to address this issue, named generalized mean distance-based k-nearest neighbor. Specifically, multi-generalized mean distances are introduced along with the nested generalized mean distance that rely on the properties of the generalized mean. Accordingly, multi-local mean vectors of a specific pattern in every group are estimated through deploying its class-specific $k$ nearest neighbors. Using the obtained $k$ local mean vectors per group, the related $k$ generalized mean distances are estimated and thereby deployed for designing the categorical nested generalized mean distance.
Similarly, in \cite{gou2019local}, the authors introduce a local mean representation-based KNN aiming at further improving the classification performance and overcoming the principal drawbacks of conventional KNN. Explicitly, they select the categorical KNN coefficients of a particular pattern to estimate the related categorical k-local mean vectors. Following, a linear combination of the categorical k-local mean vectors is used to represent the particular pattern. Moving forward, in order to capture the group of this latter, group-specific representation-based distances between the particular pattern and the categorical k-local mean vectors are then considered.

Moreover, in \cite{gou2019locality}, two locality constrained representation-based KNN rules are presented to design an improved KNN classifier. The first one is a weighted representation-based KNN rule, in which the test pattern is considered as a linear aggregation of its KNN samples from every group, while the localities of KNN samples per group are represented as the weights constraining their related representation elements. Following, a classification decision rule is used to calculate the representation-based distance between the test pattern and the group-specific KNN coefficients.
On the other side, the second rule is a weighted local mean representation-based KNN, where k-local mean vectors of KNN coefficients per group are initially estimated and then utilized to represent the test pattern. On the other hand, aiming at improving the performance of existing KNN classifiers and making them scalable and automatic, granular ball computing has been used in various frameworks. This is the case of \cite{xia2019granular}, where a granular ball KNN (GBKNN) algorithm is developed, which could perform the classification task on large-scale data sets with low computation. In addition, it provides a solution to automatically select the number $k$ of clusters.

\subsubsection{Improved k-means clustering}
In addition to the use of KNN and its variants, K-means clustering (KMC) is another important data clustering method. It has been widely investigated to classify similar data into the same cluster in large-scale data sets for different applications, such as appliance identification \cite{chui2013appliance}, anomaly detection \cite{henriques2020combining}, cancer detection \cite{saba2020recent}, and social media analysis \cite{alsayat2016social}. Despite the simplicity of KMC, its performance was not convincing in some applications. To that end, different variants have been proposed in the literature to design efficient, scalable and robust KMC classifiers. For example, in \cite{yu2018two}, to overcome the vulnerability of the conventional KMC classifier to outliers and noisy data, a tri-level k-means approach is introduced. This was possible through updating the cluster centers because data in a specific data set usually change after a period of time. Therefore, without updating the cluster centers it is not possible to accurately represent data in every cluster. While in \cite{zhang2018improved}, the authors focus on improving both the accuracy and stability of the KMC classifier. This has been achieved by proposing a k-means scheme based on density Canopy, which aims at solving the issue corresponding to the determination of the optimal number $k$ of clusters along with the optimum initial seeds. Specifically, the density Canopy has been utilized as a pre-processing step and then its feedback has been considered as the cluster number and initial clustering center of the improved KMC technique. Similarly, in \cite{lu2019improved}, an incremental KMC scheme is introduced using density estimation for improving the clustering accuracy. Explicitly, the density of input samples has been firstly estimated, where every primary cluster consists of the center points having a density superior than a given threshold along with points within a specific density range. Following, the initial cluster has been merged with reference to the distance between the two cluster centers before dividing the points without any cluster affection into clusters nearest to them.

On the other hand, in some specific data sets, e.g. real-world medical data sets, data samples could pertain to more than one cluster simultaneously while traditional KMC methods do not allow that since they are developed based on an exclusive clustering process. Therefore, an overlapping k-means clustering (OKMC) scheme is proposed in \cite{whang2015non} to overcome that issue, which have intrinsically overlapping information. Similarly, in \cite{khanmohammadi2017improved}, the authors introduce a hybrid classifier that aggregates k-harmonic means and OKMC to address the sensitivity problem of the latter to initial cluster centroids.

\subsection{Drawbacks and limitations}
Despite the fact that the outlined event-based NILM systems have recently been widely examined in the state-of-the-art, they can be affected by certain problems and limitations, which impede the development of powerful NILM architectures and even increase the difficulty to implementing real-time NILM systems. Moreover, most of these issues have not yet been overcome. For example, most of existing solutions suffer from a low disaggregation accuracy. Therefore, these approaches need deeper investigation in order to improve their performance. Moreover, they are usually built upon detecting transient states, which can limit their detection accuracy if multiple appliances are turning on/off simultaneously. In addition, most of the reviewed NILM systems are only validated on one category of data with a unique sampling frequency. This restricts the applicability of these techniques on different data repositories. On the other hand, most of the existing classifier have some issue to accurately identify appliance-level data especially if the validation data set is imbalanced.

To overcome the aforementioned limitations, we present, in this framework, a novel non-intrusive load identification, which relies on (i) shifting power fingerprints into 2D space, (ii) deriving binary characteristics at local regions, (iii) representing the extracted features in the decimal field, and (iv) going back to 1D space via capturing novel histograms of the 2D representations. Following, these steps can help in designing a robust identification approach, which has various benefits; (i) via transforming the appliance signatures into 2D space, novel appliance footprints are developed that describe each appliance fingerprint in another way and texture descriptions are derived from local regions using square kernels; (ii) the proposed strategy helps in identifying appliances at accurately without depending on the devices' states (i.e. steady or transient); (iii) the proposed scheme can support real-time applications because it can be run at a low computation cost. Specifically, it acts as a dimensionality reduction component as well, where short characteristic histograms having only 256 samples are collected at the final stage to represent every appliance, and (iv) an improved KNN algorithm has been developed to overcome the issues occurring with imbalanced data sets and improve the appliance identification performance. Moreover, our 2D descriptor can be trained via simple machine learning algorithms without the need to deploy deep leaning models, which usually have a high computation complexity.

\begin{figure}[b!]
\begin{center}
\includegraphics[width=16cm, height=10.9cm]{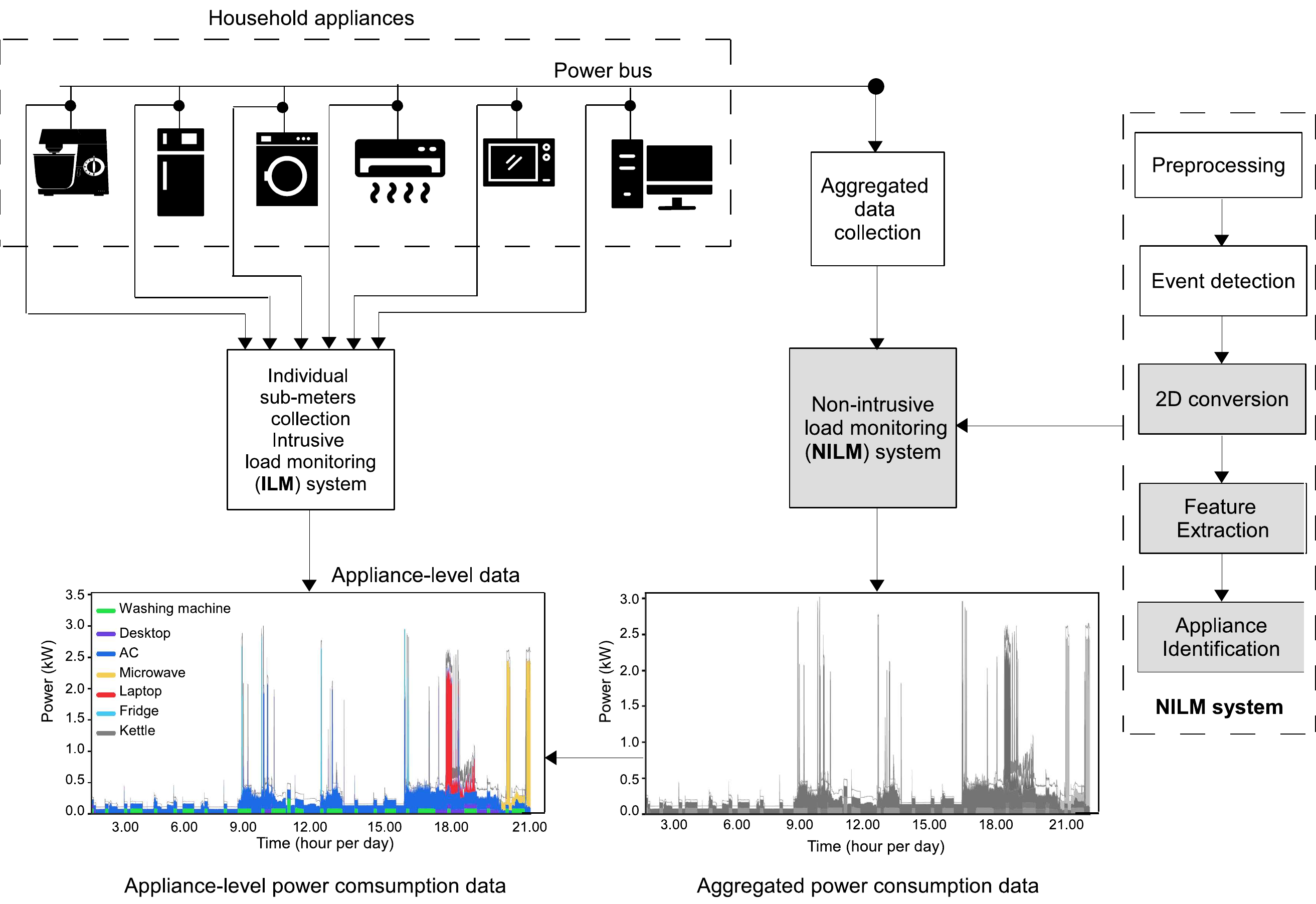}
\end{center}
\caption{Flowchart of the proposed NILM framework.}
\label{NILM-system}
\end{figure}

\section{Proposed NILM based on 2D feature extraction} \label{sec3}
This section focuses on presenting the principal steps of the proposed appliance identification system, which relies on the application of an original 2D descriptor. Accordingly, the flowchart of the proposed NILM system is portrayed in Fig. \ref{NILM-system}. It is clear that the 2D-based load identification system represents the fundamental part of the NILM system.

\subsection{Background of local 2D feature extraction}
In recent years, 2D local feature extraction schemes have received significant attention in various research topics, including image and video processing \cite{Tao2019}, breast cancer diagnosis \cite{Kumar9097394}, face identification \cite{Gong7812744} and fingerprint recognition \cite{Ramirez8681394}. They are generally deployed to derive fine-grained characteristics after partitioning the overall 2D representation into various local regions using small kernels. Explicitly, a local feature extraction can be applied at each local region  of the 2D representation to draw pertinent features about the neighborhood of each key-point. The multiple features derived from several regions are then fused into a unique, spatially augmented characteristic vector, in which the initial signals are effectively represented.

\subsection{Event detection}
For the event detection step, various event detection schemes are proposed in the state-of-the-art. Event detection techniques are split into three main groups \cite{Batra:2019:TRS}: specialized heuristics, probabilistic models and matched filters \cite{Batra:2019:DRS,Lu8090442}. In this framework, the pre-processed aggregated power is segregated into different sections using the edge detector module \cite{Batra2019} implemented in the NILMTK platform \cite{Batra2014ACM}. Accordingly, the on/off events of electrical devices are generally picked up via the analysis of power level variations in the aggregated signal. This event detector has been elected because of its simplicity and availability of its source code in the NILMTK platform.

\subsection{Local power histogramming (LPH) descriptor}
The proposed appliance identification scheme relies mainly on transforming the appliance consumption signals into 2D  space and therefore treating the appliance recognition task as a CBIR problem. With this in mind, all image descriptors could be utilized to extract the fine-grained properties of the obtained 2D power signal representations.
 
In that respect, the proposed LPH-based feature extraction scheme transforms appliance signals to image representations. Following, an examination local regions around each power sample is performed using a block partition procedure to collect local features. Explicitly, LPH descriptor is introduced for abstracting histogram-based descriptions of the 2D representations of power observations. Accordingly, LPH performs a binary encoding of power blocks through comparing the central power sample of each block with its neighbors. 

\begin{figure}[b!]
\begin{center}
\includegraphics[width=1.0\textwidth]{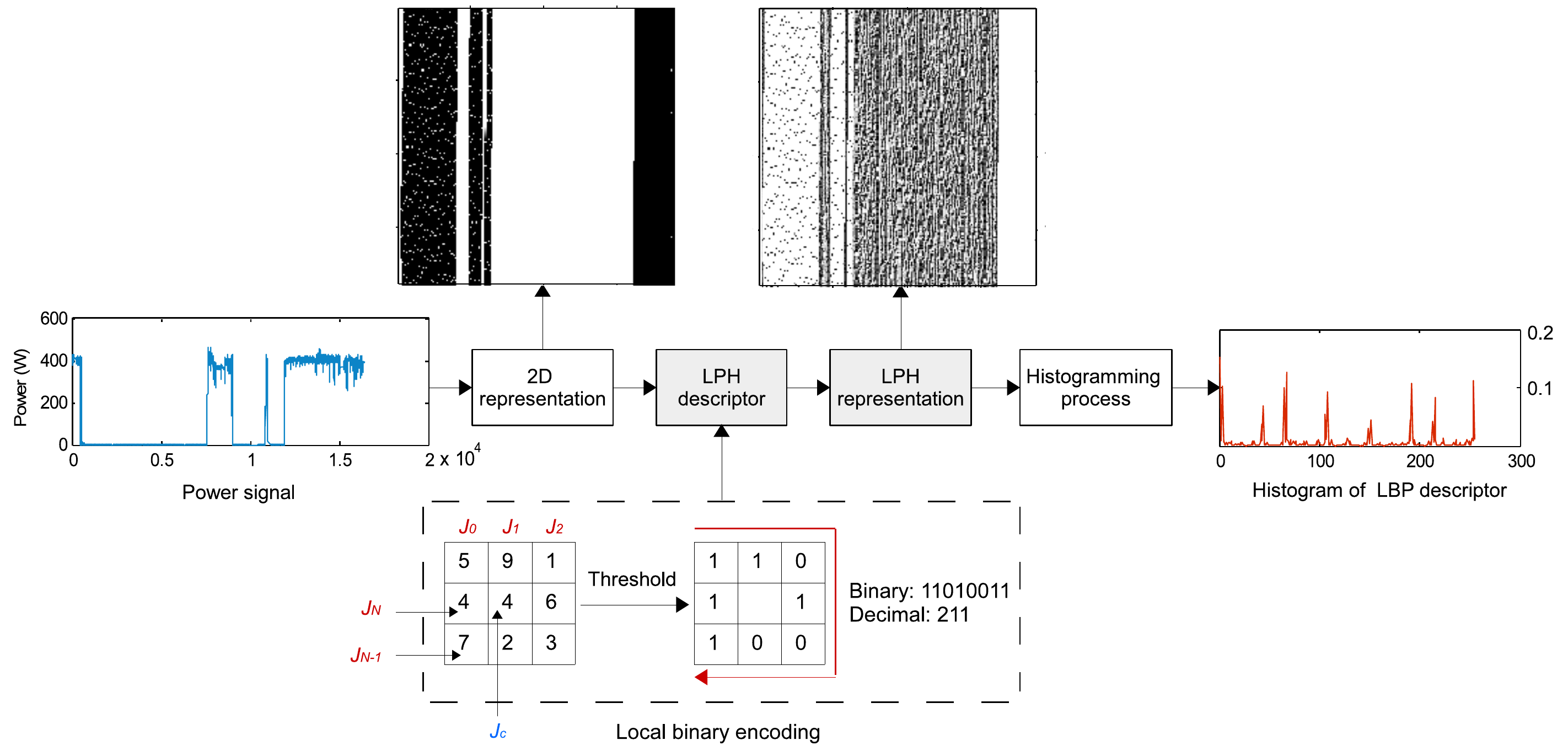}
\end{center}
\caption{Block diagram of the LPH descriptor: Example using a patch of size $3 \times 3$ (N = 8).}
\label{2D-flowchart}
\end{figure}

Fig. \ref{2D-flowchart} explains the flowchart of the proposed LPH description scheme. A comparison of each central power observation is conducted with its power neighboring in a kernel of $N \times N$ power samples through subtracting the central power value from the neighboring power patterns. Following, a binary encoding procedure is applied where the positive values of the subtractions are moved to 1, on the flip side, the negative values are considered as 0. Next, a binary sequence is then acquired by means of a clockwise-based comparison process. Consequently, the gathered binary samples represent the corresponding LPH codes. Moving forward, the overall binary sequences are gleaned from all the regions (kernels) to form a binary array, which in turn, is converted to the decimal field. Specifically, each binary sequence extracted from a specific block is converted to decimal (as it is illustrated in Fig. \ref{2D-flowchart}). Lastly, a histogramming procedure is applied on the resulted decimal array, in which an LPH histogram is extracted to represent the initial power signal. The whole steps of the proposed LPH descriptor are summarized in Algorithm \ref{algo1}.

%\RestyleAlgo{ruled}
\begin{algorithm}[t!]
\SetAlgoLined

\KwResult{ $\mathbf{B}_{LPH}$: The histogram of local power histograms (LPH) }

a. Define the array $Y(i,j)$ of the appliance power signatures, where $i$ presents the index of appliance power sequences, and $j$ stands for the index of the samples in every sequence;

 \While{$i \leq M$ (\textnormal{with} $M$ \textnormal{the total number of appliance signatures in the overall database})}{

\textbf{Step 1.} Normalize and transform the appliance signature $Y(i,:)$ into 2D space (image representation), as explained in Fig. \ref{1D-to-2D}.

\textbf{Step 2.} Calculate the LPH values of each power pattern $(u_{c},v_{c})$ in each specific kernel of size $S \times S$, by comparing the central power pattern with its neighbor as follows:
\begin{equation}
LPH_{n,S}(u_{c},v_{c})=\sum\limits_{n=1}^{N-1}b(j_{n}-j_{c})2^{n}
\end{equation}
where $j_{c}$ refers to the central power sample, $j_{n}$ represents the $n^{th}$ surrounding power neighbor in a patch of size $S \times S$ and $N = S^{2}-1$. Moving forward, a binary encoding function $b(u)$ is generated as:
\begin{equation}
b(u)=\left\{ 
\begin{array}{cc}
1 & \mathrm{if}~u \geq 0 \\ 
0 & \mathrm{if}~u < 0%
\end{array}%
\right. 
\end{equation}

\textbf{Step 3.} Glean the binary samples $LPH_{n,S}(u_{c},v_{c})$ generated from every kernel and therefore transform the obtained binary data into a decimal field in order to design a new decimal array $I_{D}$ (as it is explained in Fig. \ref{2D-flowchart}).

\textbf{Step 4.} Perform a histogramming procedure on the obtained decimal matrix for extracting an LPH histogram $H_{LPH}(n,S)$, which is measured from each patch. Thus, the resulted histogram is then used as a texture feature vector to represent the initial appliance signature. Finally, after conducting the histogramming process, a description histogram $H_{LPH}(n,S)$ is produced, which has $2^{N}$ patterns (i.e. with relation to the $2^{N}$ binary samples generated by $N$ power sample neighbors of each block of data). 
\begin{equation}
H_{LPH}(n,S) = hist(I_{D}) = [H_{1}, H_{2}, \cdots, H_{2^{N}}]
\end{equation}

\textbf{Step 5.} Normalize the resulted histogram to make the value of each bin in the range [0,1].
\begin{equation}
\mathbf{B}_{LPH}^{i}= \mathrm{Normalize}(H_{LPH}(n,S))=b_{1},b_{2},\cdots ,%
~b_{2^{N}}=\frac{H_{1}}{\textstyle\sum_{m=1}^{2^{N}}H_{1}},~\frac{%
H_{2}}{\textstyle\sum_{m=1}^{2^{N}}H_{2}},~\cdots ,~\frac{H_{2^{N}}}{%
\textstyle\sum_{m=1}^{2^{N}}H_{2^{N}}}  
\label{eq8}
\end{equation}

}
\caption{The principal steps of the proposed LPH descriptor deployed to derive LPH features from the $M$ appliance power signals.}
\label{algo1}
\end{algorithm}

Moving forward, a histogram of 256 samples is derived to represent each appliance signature, which has significantly lower number of samples compared to the initial signal. Accordingly, LPH helps also in reducing the dimensionality of the appliance power signals. Therefore, this leads to efficaciously reducing the computation cost of our NILM system.

\subsection{Improved k-nearest neighbors (IKNN)} \label{ClassModel}
This stage is responsible on predicting the labels of each power consumption observations $P(t)$ that belongs to a specific micro-moment group. Consequently, the class identification step of SAD-M2 is applied in two stages using a 10-fold validation, i.e. the training and test. In the first one, device load usage fingerprints are learned along with their labels generated based on the rule-based algorithm described previously. Accordingly, $9$ folds of the database are utilized randomly in each training phase while the remaining fold is employed for the test purpose. 

%Thereby, based on the analysis of the power consumption gleaned from a specific device and other parameters described in Algorithm \ref{algo1}, it can predict the the class of each power sample $P(t)$.
 
Moving forward, selecting the value of $K$ is of utmost importance for KNN model. However, power abnormality detection data sets suffer from the imbalanced classes issue, in which some classes include more consumption observations (i.e. majority classes) than other classes (i.e. minority classes). Accordingly, a salient drawback of conventional KNN schemes is related to the fact that if $K$ is a fixed, user-defined value, the classification output will be biased towards the majority groups in most of the application scenarios. Therefore, this results in a miss-classification problem.

To avoid the issue encountered with imbalanced data set, some works have been proposed with the aim of optimizing the value of $K$, such as \cite{Liu2011,Zhang7858565}. However, they are very complex to implement and can significantly increase the computational cost, which hinders developing real-time abnormality detection solutions. In contrast, in this paper, we introduce a simple yet effective improvement of KNN, which can maintain a low computational cost. It is applied as explained in Algorithm \ref{improvedKNN}.

\begin{figure}[t!]
\begin{center}
\includegraphics[width=1.0\textwidth]{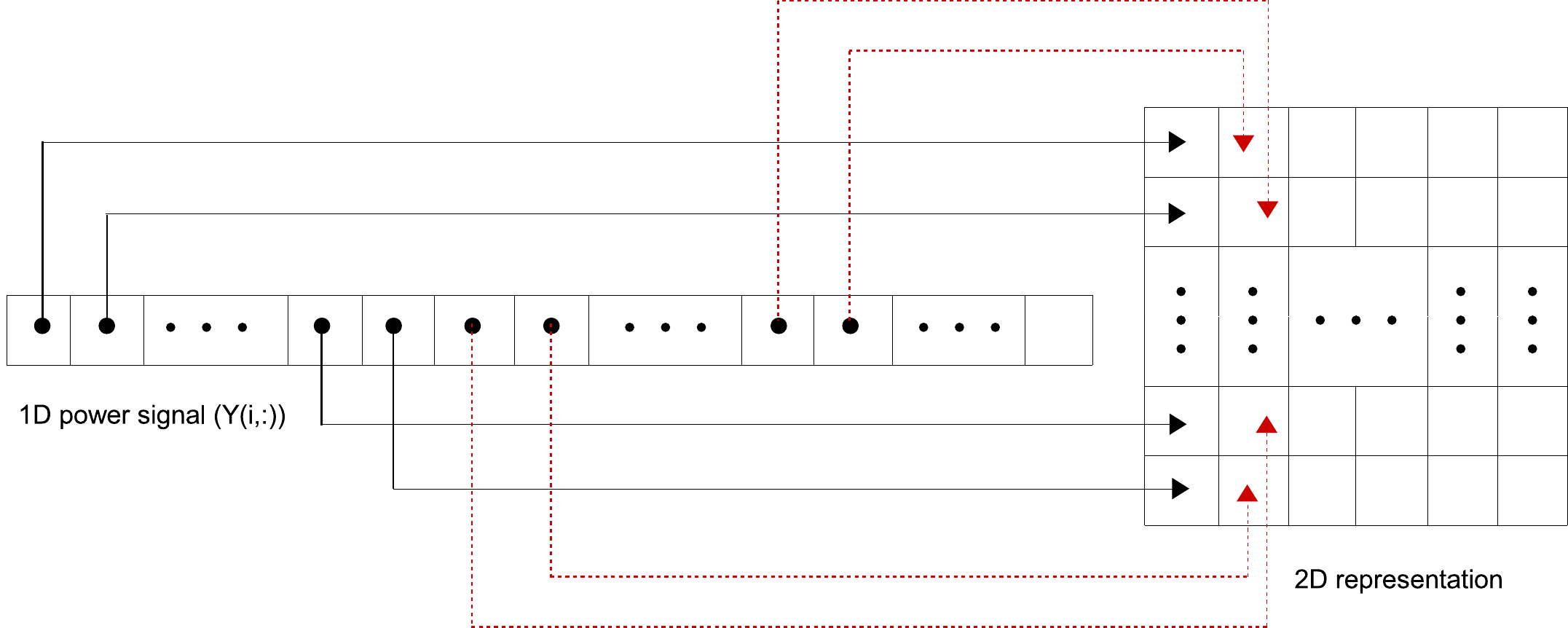}
\end{center}
\caption{Conversion of 1D signal into 2D representation.}
\label{1D-to-2D}
\end{figure}

\begin{algorithm}[t!]
\SetAlgoLined
\KwResult{Predicting class labels of test samples}
Read the training appliance histograms extracted using LPH in Algorithm \ref{algo1}. \\

\While{$j \leq J$ (\textnormal{with} $J$ \textnormal{is the number of test appliance histograms to be identified})}{

\textbf{Step 1:} Compute the information-entropy of every appliance histogram $b$ that is deployed to estimate its information gain. Thus, it operates as the weight of appliance histograms power consumption observations to allocate priorities to each of them;

\begin{equation}
E(B)=-\sum\limits_{i=1}^{n}a_{i}\log _{2}(a_{i})
\end{equation}

where  $B$ is the training ensemble, $\left\vert B\right\vert $ as the
number of training data, $a_{i}=\left\vert c_{i},B\right\vert $ $/$ $%
\left\vert B\right\vert $ and $a_{i}$ refers to the probability that an random
histogram in $B$ pertains to class $c_{i}$ 

\textbf{Step 2:} Define the $k$ values of the training ensemble;

\textbf{Step 3:} Partition the training ensemble into $m$ sub-groups;

\textbf{Step 4:} Estimate the mean value of each sub-group to derive its center;

\textbf{Step 5:} Identify the sub-group that is closest to a test histogram $b_{j}$ via estimating the Euclidean distance between each test observation and
the center of each sub-group as follows:

\begin{equation}
d_{j}(b_{c_{i}},b_{j})=\sqrt{(b_{c_{i}})-(b_{j})^{2}}
\end{equation}
where $c_{i}$ represents the central instance of the $i^{th}$ sub-group and $i=1,2,\cdots ,m$.

\textbf{Step 6:} Estimate the weighted-Euclidean distance $wd_{j}$ between the test histogram $b_{j}$ and every histogram in the closest sub-group as follows: 

\begin{equation}
wd_{j}(b_{i},b_{j})=\sqrt{w_{i}(b_{i}-b_{j})^{2}}
\end{equation}

Therefore, this results in determining the $k$ nearest neighbors;

\textbf{Step 7:} Compute the weighted class probability of the test histogram $b_{j}$ as follows:
\begin{equation}
c(b_{j})=\arg \underset{c\in C}{\max }\sum\limits_{i}^{k}w_{i}\delta
(c,c(y_{i})) ~~  \label{eq5}
\end{equation}
where $y_{1},y_{2},\cdots ,y_{k}$ refer to the $k$ nearest neighbors of the test histogram $b_{j}$, $C$ denotes the finite set of the appliance class labels, $\delta (c,c(y_{i}))$ $=1$ if $c=c(y_{i})$ and $\delta (c,c(y_{i}))$ $=0$ otherwise. 

}

\caption{IKNN algorithm used to classify appliances based on their LPH signatures.}
\label{improvedKNN}
\end{algorithm}

%\end{itemize}

Overall, the proposed improved KNN helps in improving the appliance identification performance through enhancing the classification accuracy and F1 score results in addition to reducing the execution time as it will be demonstrated in the next step. Therefore, this could help in developing real-time abnormality detection solutions.

\section{Evaluation and discussion} \label{sec4}
We concentrate in this section on presenting  the outcomes of an extensive empirical evaluation conducted on four real-world data sets, namely UK-DALE \cite{UK-DALE2015}, GREEND \cite{GREEND2014}, PLAID \cite{PLAID2014} and WHITED \cite{WHITED2016}. They are  which are vastly deployed to validate  NILM and load identification frameworks in the state-of-the-art. 

\subsection{data set description}
The three power consumption repositories considered in this framework are gleaned at distinct sampling rates (i.e. 1/6 Hz, 30 kHz and 44 kHz) to perform a thorough evaluation study and inform the effectiveness of the proposed solution when the sampling rate of the recorded appliance consumption signals varies.

Under UK-Dale, power usage footprints have been gathered for a long time period ranging 2 to 4 years at both sampling frequencies of 1/6 Hz and 16 kHz (for aggregated data). In order to assess the performance of proposed scheme, we exploit the consumption fingerprints gleaned from a specific household at 1/6 Hz, which encompasses nine appliance categories and each category includes a large number of daily consumption signatures. Moving forward, power traces of six different appliances collected under GREEND \cite{GREEND2014} are also considered, in which a sampling frequency of 1 Hz has been used to record energy consumption footprints for a period of more than six months. Under PLAID, the power signatures of 11 device groups have been recorded on the basis of a frequency resolution of 30 kHz. Moreover, load usage footprints of the WHITED have been gleaned with reference to 11 appliance classes at a frequency resolution of 44 kHz. The properties of each data set, their appliance categories and the number of observed appliance/days are recapitulated in Table \ref{WHITED}.

\begin{table}[t!]
\caption{Properties of power consumption data sets considered in this framework, i.e. appliance classes and their number for both PLAID and WHITED, and appliance classes and number of observed days for both UK-DALE and GREEND.}
\label{WHITED}
\begin{center}

\begin{tabular}{lll|lll|lll|lll}
\hline
\multicolumn{3}{c|}{\small UK-DALE} & \multicolumn{3}{|c|}{GREEND}
& \multicolumn{3}{|c|}{\small PLAID} & \multicolumn{3}{|c}{\small WHITED} \\ 
\hline
{\small \#} & {\small Device} & {\small \#} & \# & Device
& \multicolumn{1}{l|}{\#} & {\small \#} & {\small Device} & {\small %
\# } & {\small \#} & {\small Device} & {\small \#} \\ 
& {\small class} & {\small days} &  & class & \multicolumn{1}{l|}{%
days} &  & {\small class} & {\small app} &  & {\small class} & 
{\small app} \\ \hline
{\small 1} & {\small Dishwasher} & {\small 183} & 1 & Coffee & 242 & {\small 1} & {\small Fluorescent lamp} & {\small 90} & 
{\small 1} & {\small Modems/receivers} & \multicolumn{1}{r}{\small 20} \\ 
{\small 2} & {\small Refrigerator} & {\small 214} &  & machine &  & 
{\small 2} & {\small Fridge} & {\small 30} & {\small 2} & {\small Compact
fluorescent} & \multicolumn{1}{r}{\small 20} \\ 
{\small 3} & {\small Washing machine} & {\small 210} & 2 & Radio & 242 & {\small 3} & {\small Hairdryer} & {\small 96} &  & 
{\small lamp} & \multicolumn{1}{r}{} \\ 
{\small 4} & {\small Microwave} & {\small 171} & 3 & Fridge
& 240 & {\small 4} & {\small Microwave} & {\small 94} & {\small 3}
& {\small Charger} & \multicolumn{1}{r}{\small 30} \\ 
{\small 5} & {\small Stove} & {\small 193} &  & w/freezer &  & 
{\small 5} & {\small Air conditioner} & {\small 51} & {\small 4} & {\small %
Coffee machine} & \multicolumn{1}{r}{\small 20} \\ 
{\small 6} & {\small Oven} & {\small 188} & 4 & Dishwasher & 242 & {\small 6} & {\small Laptop} & {\small 107} & {\small 5} & 
{\small Drilling machine} & \multicolumn{1}{r}{\small 20} \\ 
{\small 7} & {\small Washer/dryer} & {\small 216} & 5 & Kitchen & 242 & {\small 7} & {\small Vacuum cleaner} & {\small 8}
& {\small 6} & {\small Fan} & \multicolumn{1}{r}{\small 30} \\ 
{\small 8} & {\small Air conditioner} & {\small 157} &  & lamp &  & 
{\small 8} & {\small Incadescent light bulb} & {\small 79} & {\small 7} & 
{\small Flat iron} & \multicolumn{1}{r}{\small 20} \\ 
{\small 9} & {\small LED light} & {\small 172} & 6 & TV & 
242 & {\small 9} & {\small Fan} & {\small 96} & {\small 8} & 
{\small LED light} & \multicolumn{1}{r}{\small 20} \\ 
&  &  &  &  &  & {\small 10} & {\small Washing machine} & {\small 22} & 
{\small 9} & {\small Kettles} & \multicolumn{1}{r}{\small 20} \\ 
&  &  &  &  &  & {\small 11} & {\small Heater} & {\small 30} & {\small 10} & 
{\small Microwave} & \multicolumn{1}{r}{\small 20} \\ 
&  &  &  &  &  &  &  &  & {\small 11} & {\small Iron} & \multicolumn{1}{r}%
{\small 20} \\ \hline
\end{tabular}

\end{center}
\end{table}

\subsection{Evaluation metrics}
Aiming at evaluating the quality of the proposed appliance identification objectively, various metric are considered, including the accuracy and F1 score, normalized cross-correlation (NCC) and histogram length. The accuracy is introduced to measure the ratio of  successfully recognized devices in the testbed, but it is nonetheless not enough to evaluate the performance of an appliance identification system giving that alone it is not regarded as a reliable measure. This is mainly the the case of imbalanced data sets, in which the power samples are not uniformly distributed (e.g. in this framework, both PLAID and WHITED data sets are imbalanced). To reinforce the objectivity of the evaluation study, F1 score is also recorded as well, which is considered as a fairly trustworthy metric in such scenarios. Explicitly, F1 score describes the specified as the harmonic average of both the precision and recall measures.

\begin{equation}
Accuracy=\frac{TP+TN}{TP+FP+TN+FN}
\end{equation}
where $TP$, $TN$, $FP$ and $FN$ depict the number of true positives, true negatives, false positives and false negatives, respectively. 
\begin{equation}
F1 score = 2\times \frac{precision \times recall}{precision + recall}
\end{equation}
where $[precision=\frac{TP}{TP+TF}]$ and $[recall=\frac{TP}{TP+FP}]$.

Additionally, normalized cross-correlation (NCC) has been deployed to measure the similarity of the raw appliance signatures and LPH histograms derived form original power signals. NCC is also described via calculating the cosine of the angle $\theta$ between two power signals (or extracted characteristic histograms) $x$ and $y$:

\begin{equation}
NCC=Cos(\theta )=\frac{x\cdot y}{\left\vert x\right\vert \left\vert
y\right\vert }=\frac{\sum\nolimits_{i}x_{i}\cdot y_{i}}{\sqrt{%
\sum\nolimits_{i}x_{i}}\sqrt{\sum\nolimits_{i}y_{i}}}, ~~~~-1\leq NCC\leq 1
\end{equation}

\subsection{Performance in terms of the NCC}
It is of utmost importance to comprehend at the outset how LPH histograms varies from the initial appliance power signatures. Accordingly, this subsection focuses on investigating the nature of the relation between appliance power signatures that pertain to the same appliance class. In addition, this can aid in understanding the way LPH histograms could improve the discrimination ability between appliances belonging to different classes and on the other flip increasing the similarity ratio between appliance from the same group.  

\begin{figure}[t!]
\begin{center}
\includegraphics[width=6.5cm, height=4.2cm]{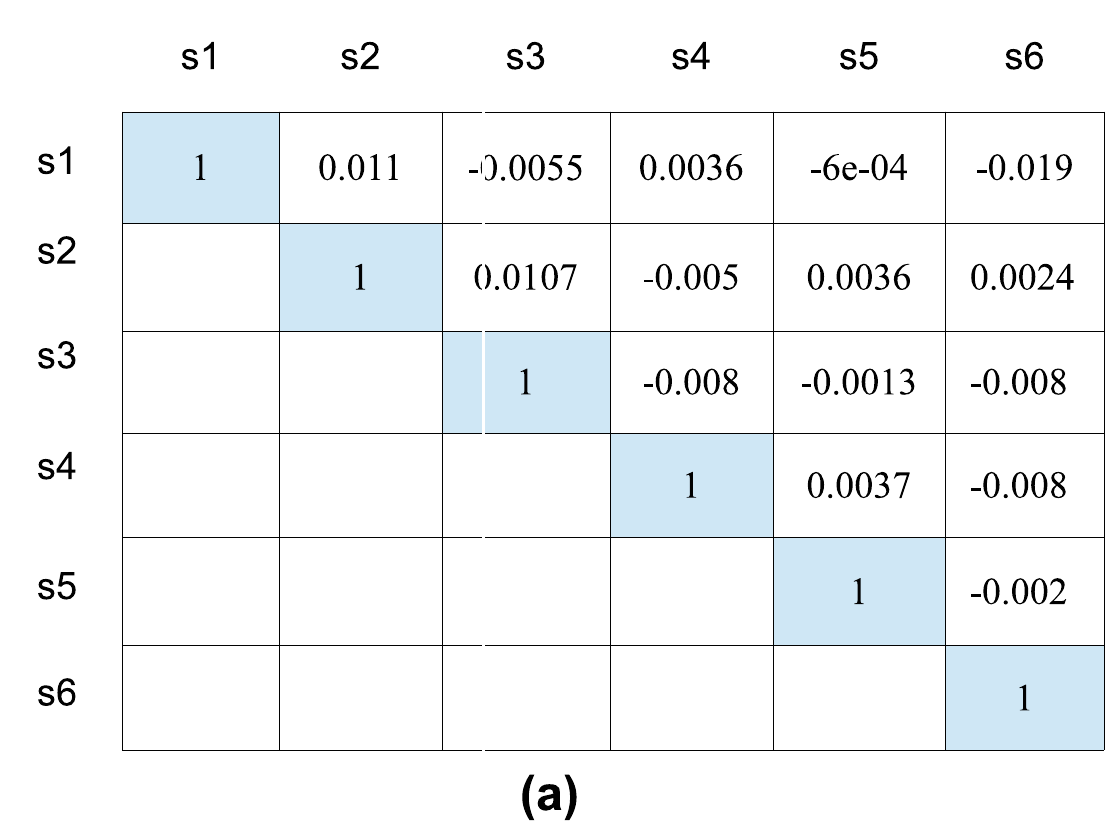}
\includegraphics[width=6.5cm, height=4.2cm]{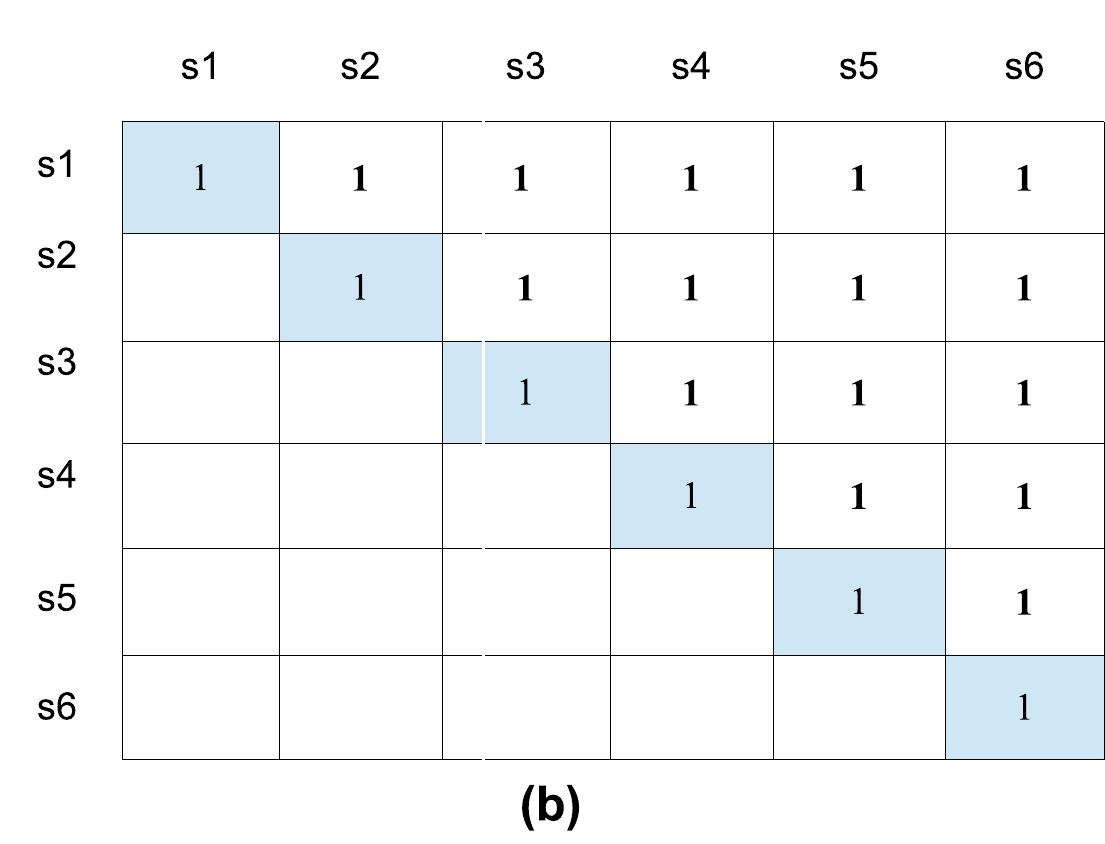}\\
(I) Coffee machine\\
\includegraphics[width=6.5cm, height=4.2cm]{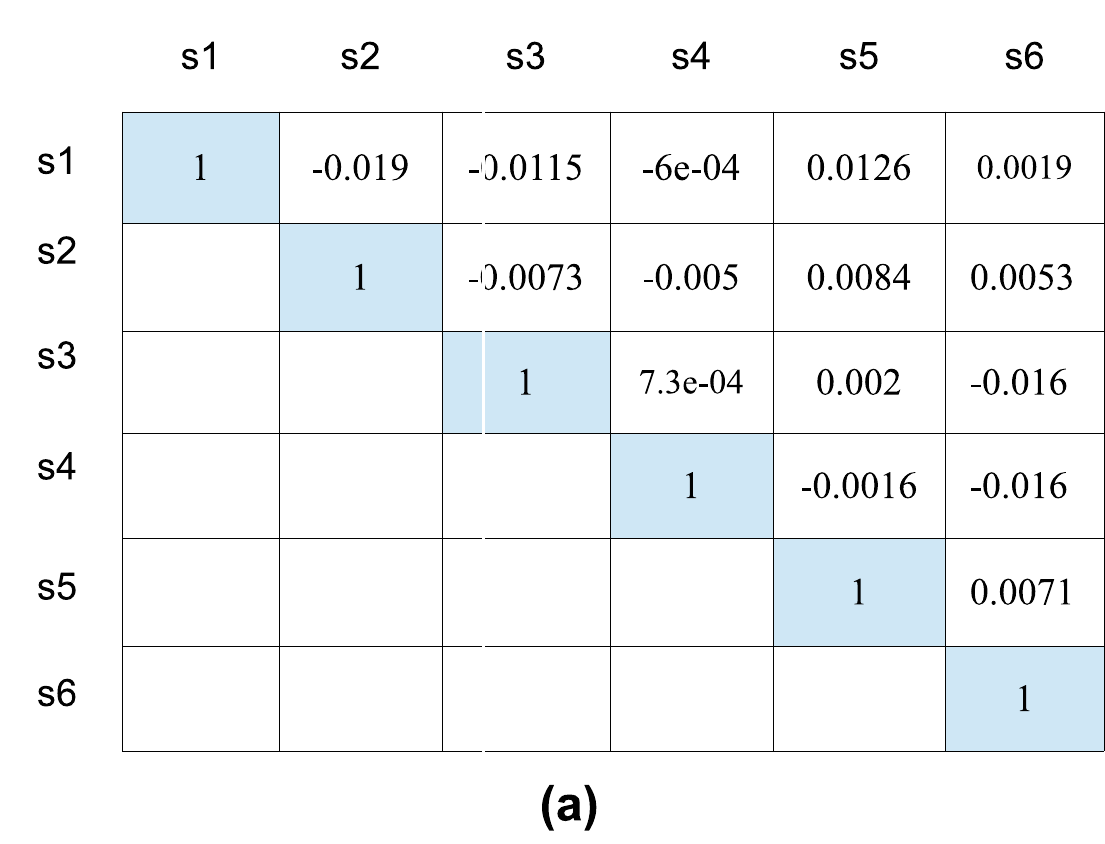}
\includegraphics[width=6.5cm, height=4.2cm]{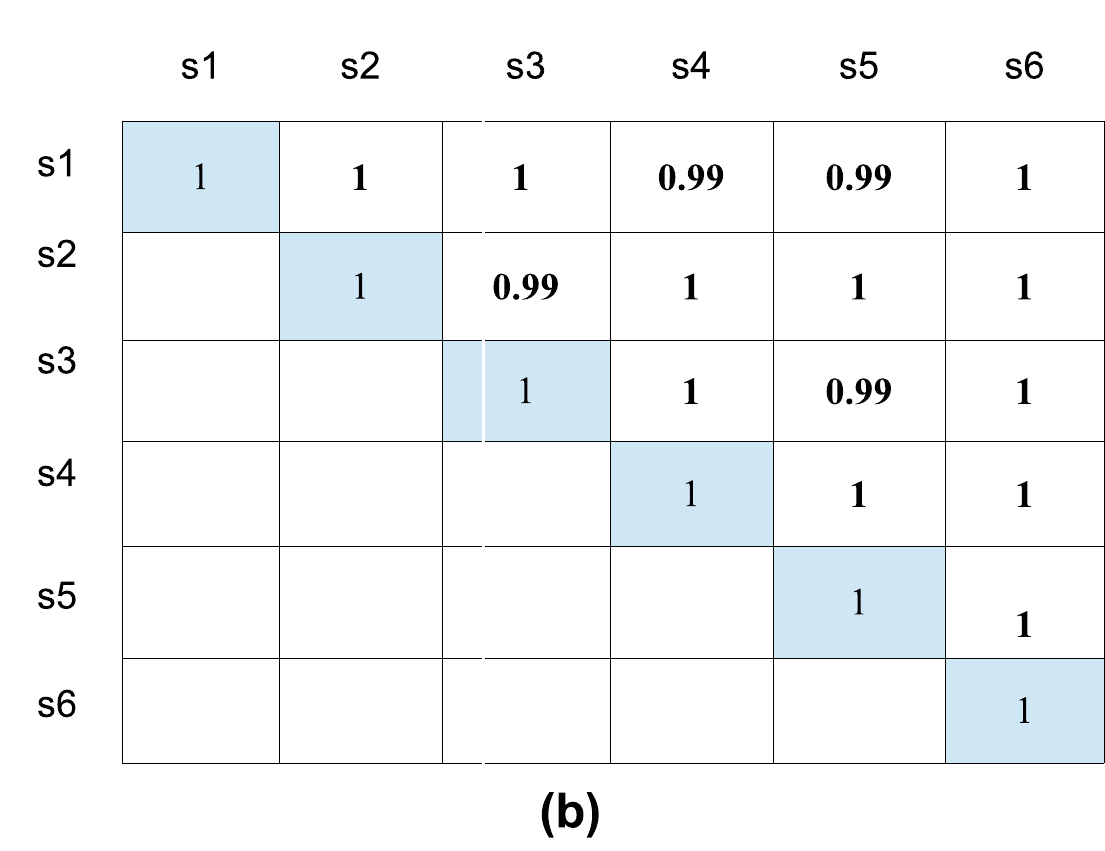}\\
(II) Radio \\
\includegraphics[width=6.5cm, height=4.2cm]{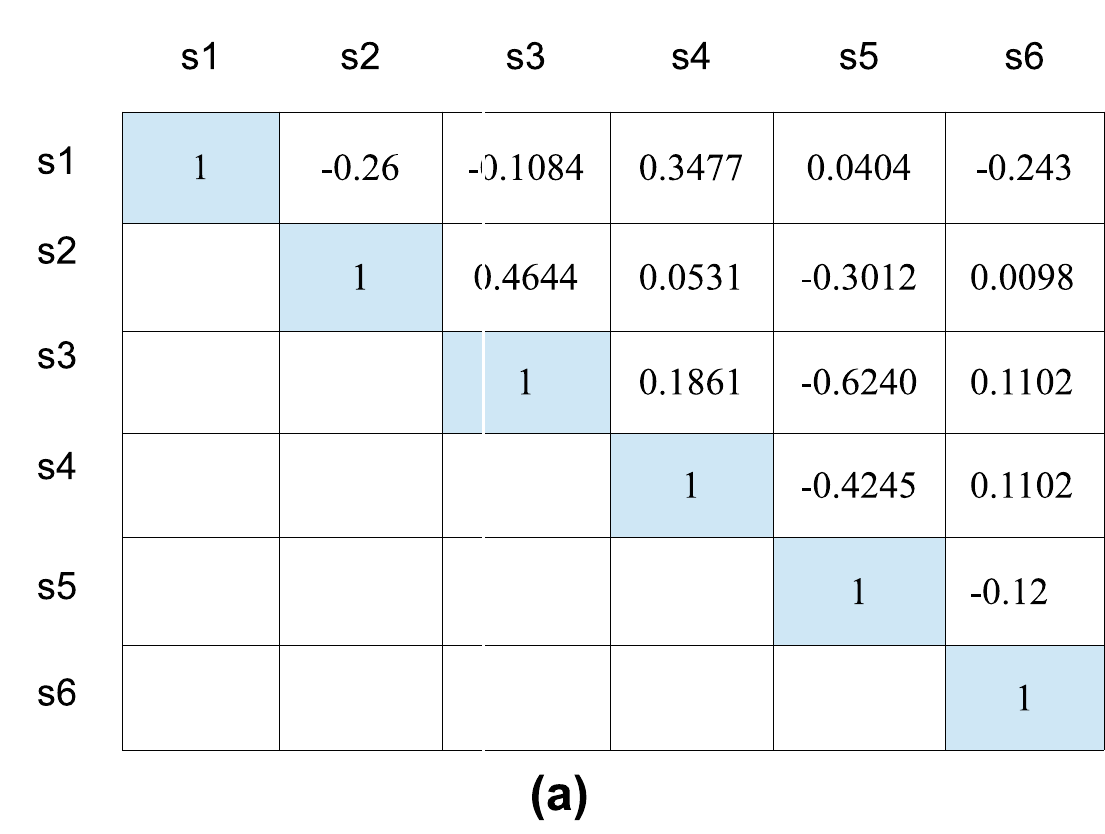}
\includegraphics[width=6.5cm, height=4.2cm]{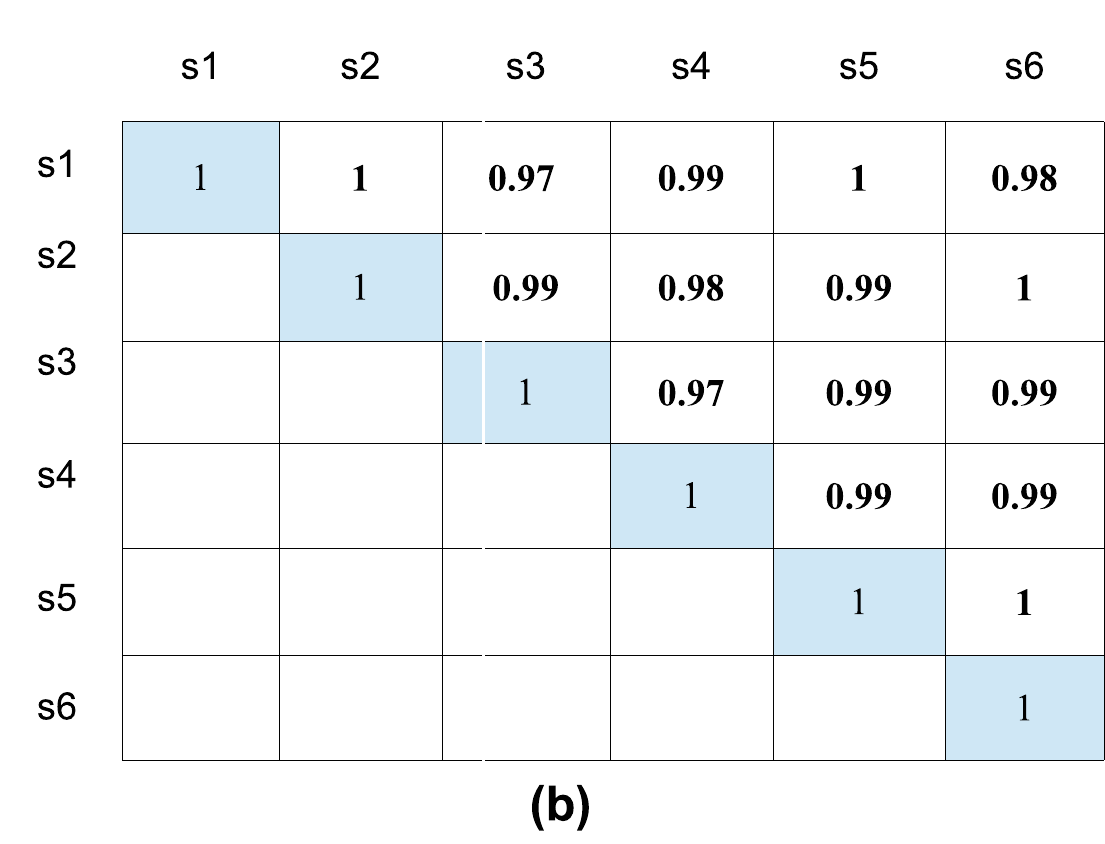}\\
(III) Fridge w/ freezer \\
\includegraphics[width=6.5cm, height=4.2cm]{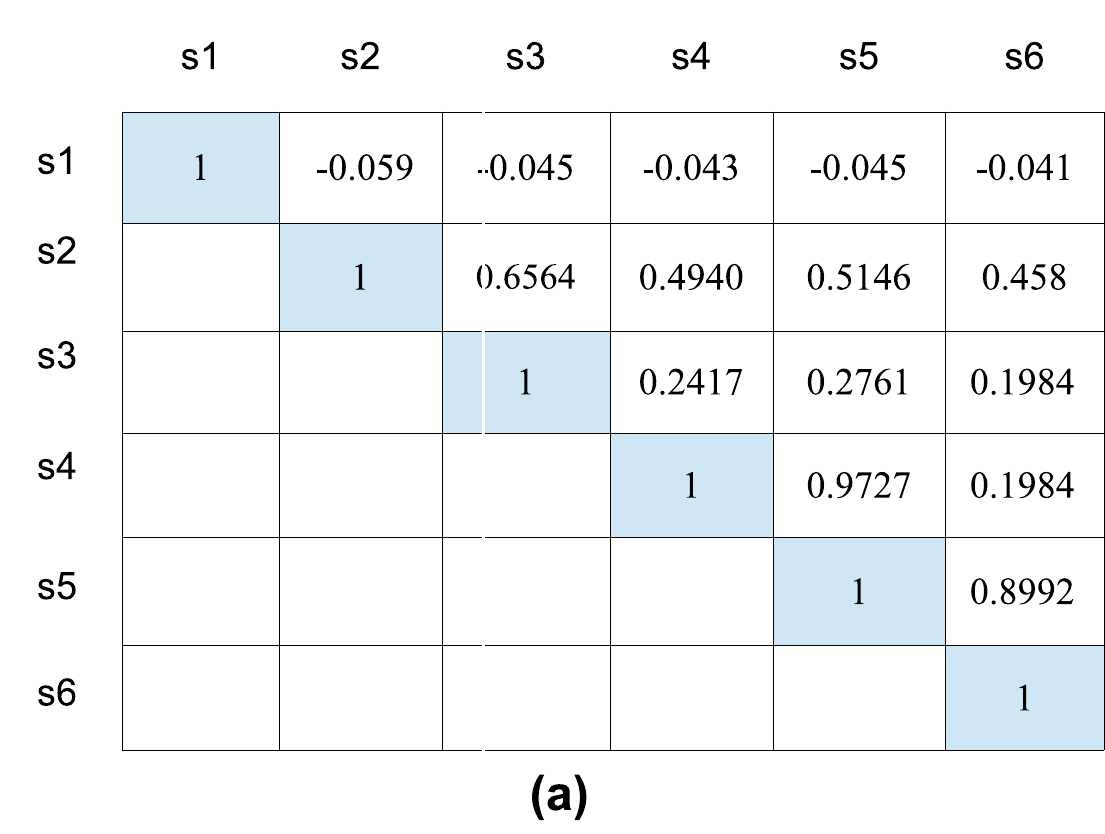}
\includegraphics[width=6.5cm, height=4.2cm]{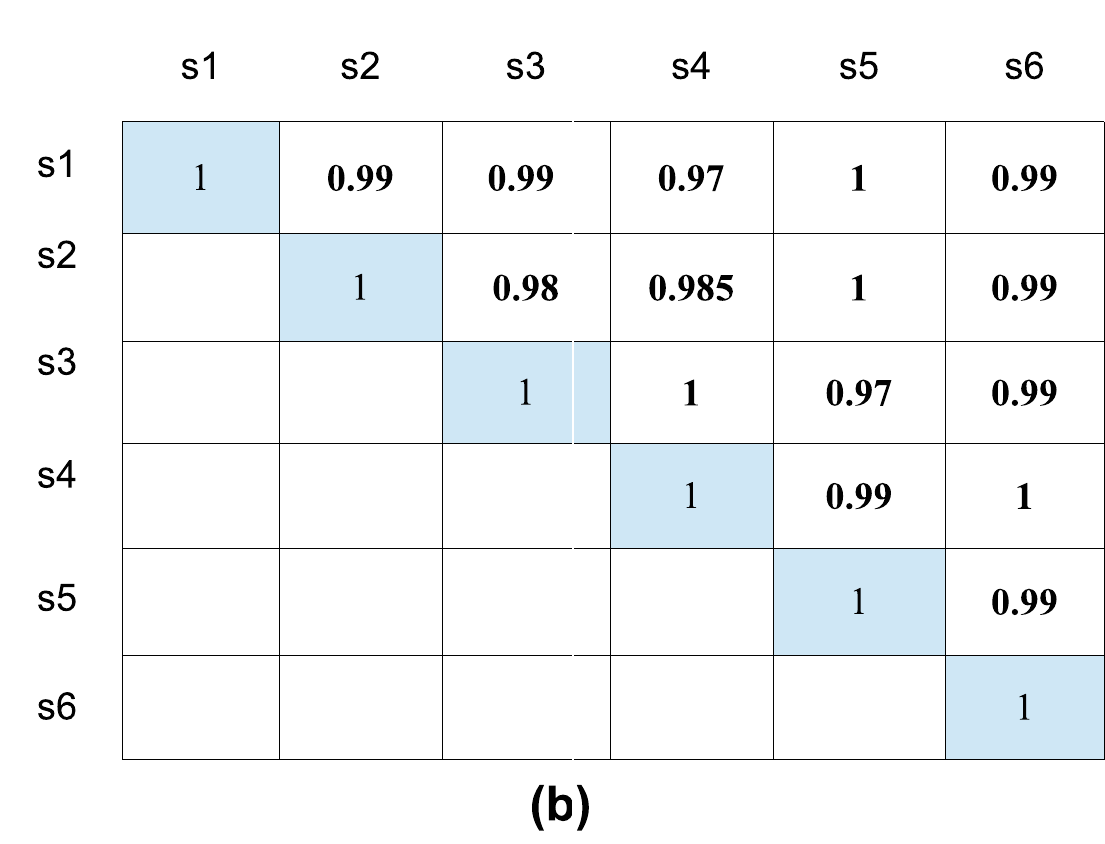}\\
(IV) Washing machine 

\end{center}
\caption{Correlation arrays computed for: (a) raw appliance signatures belonging to the same appliance category and (b) their LPH histograms.}
\label{CorrMat}
\end{figure}

To that end, six appliance signatures $s1, s2, \cdots , s6$ have been considered randomly from every device category of the UK-DALE data set. Moving forward, the NCC performance has been measured between these signatures to evidently demonstrate why the LPH can results in a better correlation between the signatures of the same device category. Fig. \ref{CorrMat} outlines obtained NCC matrices, which are calculated between the six raw power signals (left side) and the LPH feature vectors (right side), respectively. Both raw power signals and LPH vectors are gleaned from four device groups, defined as the washing machine, fridge w/ freezer, coffee machine and radio. It can be shown from the plots in the left side of Fig. \ref{CorrMat} that NCC rates are quiet low and vary randomly. Specifically, it is hard to identify a certain interval specifying the limits of the NCC rates. On the other side, when measuring the correlation between LPH vectors as indicated in the right side of Fig. \ref{CorrMat}, NCC values outperform those obtained from the raw power signals. Overall, NCC rates gleaned from LPH vectors are generally more than 0.97 for all appliance groups investigated in this correlation study.

\begin{figure*}[t!]
\begin{center}
\includegraphics[width=12.9cm, height=4.3cm]{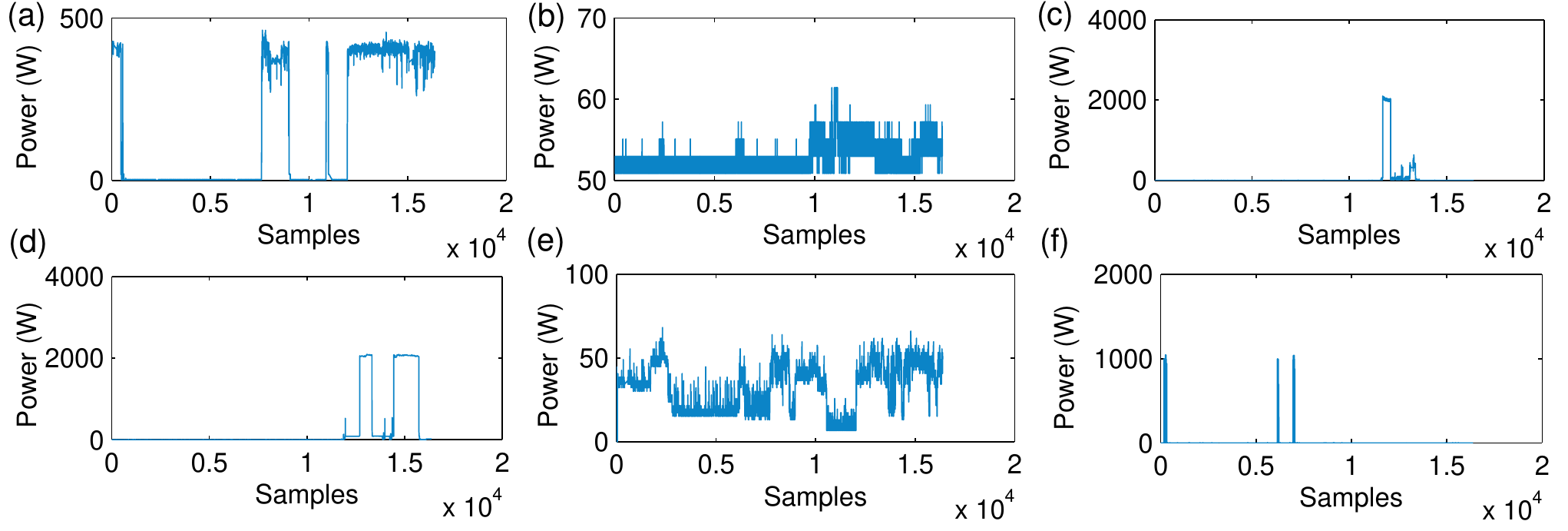}\\
I. Example of power signals from the UK-DALE data set.
\includegraphics[width=14.5cm, height=6.9cm]{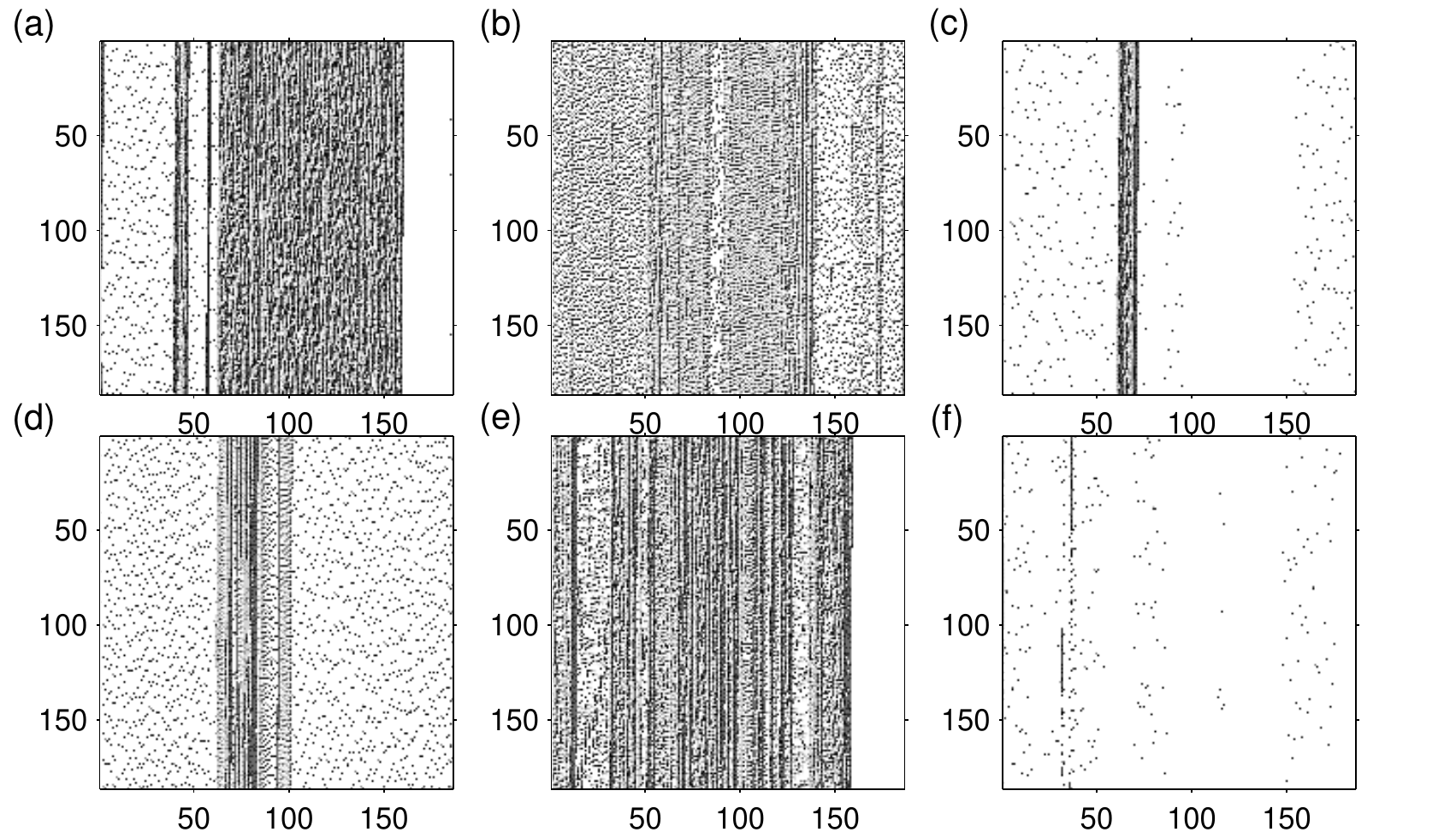}\\  
II. Image representation of LPH encoding of the power signals.
\includegraphics[width=12.9cm, height=4.3cm]{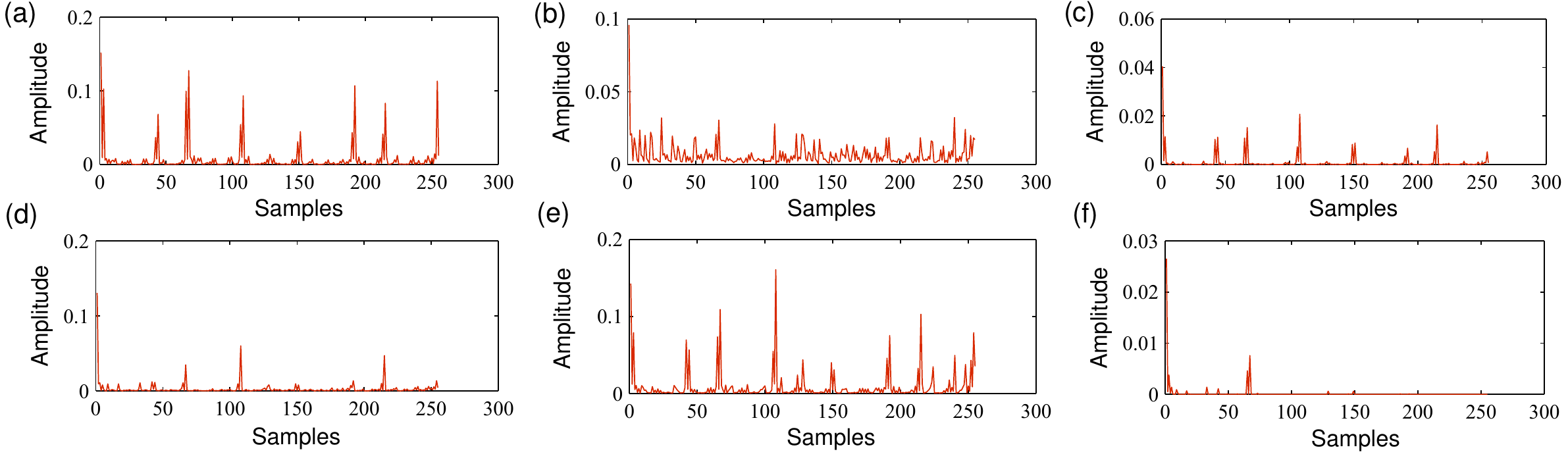}\\  
III. Histograms of LPH representations of the power signals.
\end{center}
\caption{Example of appliance signals, their 2D LPH representations and their LPH histograms from the UK-DALE data set: (a) television, (b) Network Attached Storage (NAS), (c) washing machine, (d) dishwasher, (e) notebook and (f) coffee machine.}
\label{HLPH}
\end{figure*}

Fig. \ref{HLPH} portrays an example of six appliance signals extracted from UK-DALE database, their encoded 2D representations and final histograms collected using the LPH descriptor. It is worth noting that each appliance signal is represented by a unique image in 2D space and further by a specific histogram in the final step. 
It has been clearly seen that via transforming the appliance signals into 2D space, they have been considered as image, where we can use any 2D feature extraction scheme to collected pertinent features. Moreover, through adopting the 2D representation, every power sample has been encircled by eight neighboring samples instead of only two neighbors in the 1D space. Therefore, more opportunities have been available for extracting fine-grained characteristics from each device signature in reliable way. Consequently, it could help effectively correlate between devices that pertain to the same device group, and in contrast, it increases the distance between devices corresponding  to distinct categories. In addition, the LPH-based load identification system does not relies on capturing the appliances' states (steady or transient). This represents and additional benefit of the proposed solution, which could recognize each electrical device without the need to collection state information.

Moreover, even the neighborhood is temporary distant in 2D space but this gives us various possibilities to encode the power signal. Therefore, this results in better correlation and discrimination abilities and hence a better classification performance of the power signals. In the contrary, if the signal is processed in the original 1D space, the possibilities for encoding the signal are very limited. Thus, the correlation and discrimination abilities lose their efficacy since the classifiers frequently make confusion in identifying appliances using the features generated in this space.

\subsection{Performance compared to different ML classifiers }
We present in this subsection the performance of the proposed appliance identification system based LPH-IKNN  in comparison with different classifiers, namely conventional KNN, DT, SVM, DNN and EBT. Specifically, Table \ref{ACCFscore} reports the accuracy and F1 scores collected under UK-DALE, PLAID and WHITED data set, in which a 10-fold cross validation is adopted. 

It is clearly shown that the proposed IKNN classifier based on both Euclidean distance and weighted Euclidean distance outperforms the other classification models, it provides the best results on the three data sets considered in this framework. For instance, it achieves 98.50\% and 98.49 F1 score under UK-DALE while the performance has slightly propped for both the PLAID and WHITED data sets. Accordingly, 96.85\% accuracy and 96.18\% F1 score have been obtained under PLAID and 96.5\% and 96.04\% have been attained under WHITED.
This might be justified by the rise of the sampling frequency in both PLAID and WHITED data sets, where data have been gleaned at 30 kHz and 44 kHz, respectively, in contrast to UK-DALE, in which consumption footprints have been gathered at a resolution of 1 Hz. Moreover, it is important to notice that the LPH descriptor serves not only as a feature extraction descriptor but as a dimensionality reduction approach as well. Explicitly, for each appliance signal, the resulting LPH vector encompasses only 256 samples while the initial appliance signals include much higher samples (e.g. 22491,  57600 and 30000 samples WHITED, UK-DALE and PLAID, respectively).
This drives us to determine that the proposed LPH-IKNN solution operates better under low frequencies. All in all, the performance obtained with LPH-IKNN are very promising because they are all superior than 96\% for all the data sets considered in this study.

\begin{table}[t!]
\caption{Performance of the proposed appliance identification system using the LPH descriptor in terms of the accuracy and F1 score with reference to various classifiers.}
\label{ACCFscore}
\begin{center}

\begin{tabular}{lccccccccc}
\hline
{\small Classifier } & {\small Classifier} & \multicolumn{2}{c}{\small %
UK-DALE} & \multicolumn{2}{c}{GREEND} & \multicolumn{2}{c}{\small %
PLAID} & \multicolumn{2}{c}{\small WHITED} \\ \cline{3-10}\cline{3-9}
& {\small \ parameters} & {\small Accuracy} & {\small F1 score} & Accuracy & F1 score & {\small Accuracy} & {\small F1 score} & 
{\small Accuracy} & {\small F1 score} \\ \hline
{\small LDA} & {\small /} & {\small 93.71} & {\small 93.53} & 94.55
& 94.37 & {\small 84.71} & {\small 77.93} & {\small 82.50} & 
{\small 77.41} \\ 
{\small DT} & {\small Fine, 100 splits} & {\small 97.42} & {\small 97.37} & 
97.81 & 97.69 & {\small 75.42} & {\small 66.9} & {\small %
92.5} & {\small 90.49} \\ 
{\small DT} & {\small Medium, 20 splits} & {\small 96.51} & {\small 96.5} & 
96.77 & 96.70 & {\small 65.85} & {\small 50.20} & {\small %
91.25} & {\small 90.84} \\ 
{\small DT} & {\small Coarse, 4 splits} & {\small 73.86} & {\small 69.38} & 
75.11 & 71.36 & {\small 49} & {\small 31.15} & {\small 34.16%
} & {\small 28.36} \\ 
{\small DNNs} & {\small 50 hidden layers} & {\small 71.69} & {\small 69.82}
& 74.3 & 72.42 & {\small 78.14} & {\small 76.09} & {\small %
82.37} & {\small 81.86} \\ 
{\small EBT} & {\small 30 learners, 42 k} & {\small 82.51} & {\small 81.26}
& 84.66 & 82.71 & {\small 82.57} & {\small 74.98} & 
{\small 91.66} & {\small 88.67} \\ 
& {\small splits} &  &  &  &  &  &  &  &  \\ 
{\small SVM} & {\small Linear Kernel} & {\small 94.84} & {\small 95} & 
95.39 & 95.48 & {\small 81.85} & {\small 71.61} & {\small %
84.58} & {\small 82.52} \\ 
{\small SVM} & {\small \ Gaussian kernel} & {\small 89.31} & {\small 98.93}
& 90.61 & 90.05 & {\small 85} & {\small 77.57} & {\small %
84.91} & {\small 87.91} \\ 
{\small SVM} & {\small Quadratic kernel} & {\small 93.93} & {\small 93.81} & 
94.72 & 94.13 & {\small 89.14} & {\small 85.34} & {\small %
92.5} & {\small 89.07} \\ 
{\small KNN} & {\small k=10/Weighted} & {\small 96.96} & {\small 96.81} & 
97.23 & 96.97 & {\small 82.14} & {\small 73.57} & {\small %
87.91} & {\small 82.71} \\ 
& {\small \ Euclidean dist} &  &  &  &  &  &  &  &  \\ 
{\small KNN} & {\small k=10/Cosine dist} & {\small 96.13} & {\small 96.01} & 
96.79 & 96.55 & {\small 75.57} & {\small 65.57} & {\small %
84.58} & {\small 80.1} \\ 
{\small KNN} & {\small k=1/Euclidean \ dist} & {\small 97.45} & {\small 97.41%
} & 97.60 & 97.36 & {\small 91.75} & {\small 89.07} & 
{\small 92.43} & {\small 89.97} \\ 
{\small IKNN} & {\small k=5/Weight Euclidean } & {\small \textbf{98.50}} & {\small %
\textbf{98.49}} & \textbf{98.84} & \textbf{98.77} & {\small \textbf{96.85}} & {\small \textbf{96.48}} & 
{\small \textbf{96.55}} & {\small \textbf{96.34}} \\ 
& {\small \ distance + Euclidean} &  &  &  &  &  &  &  &  \\ 
& {\small \ \ distance } &  &  &  &  &  &  &  &  \\ \hline
\end{tabular}

\end{center}
\end{table}

On the other side, it is worth noting that the proposed LPH descriptor can be trained using simple ML algorithms without the need to deploy deep leaning models, which usually have a high computation complexity. In this direction, it was obvious that conventional classifiers, e.g. LDA, DT, EBT, SVM and KNN outperforms significantly the DNN classifier, especially under UK-DALE and GREEND data sets.

\subsection{Comparison with existing 2D descriptors}
The promising results of the proposed LPH obtained under the three data sets considered in this study has pushed us to investigate the performance of other 2D descriptors in comparison with our solution. Accordingly, in this section, we investigate the performance of three other feature extraction schemes.

\begin{itemize}
\item \textit{Local directional patterns (LDP)}: After transforming the power signal into 2D space, for each pattern of the power array, an 8-bit binary sequence is derived using LDP \cite{Perumal2016}. The latter is measured via the convolution of small kernels from the power array (e.g. $3 \times 3$) with the Kirsch blocks in 8 different orientations. Fig. \ref{kenels} portrays an example of the Kirsch blocks used in LDP. 

\begin{figure}[t!]
\begin{center}
\includegraphics[width=9.5cm, height=4.9cm]{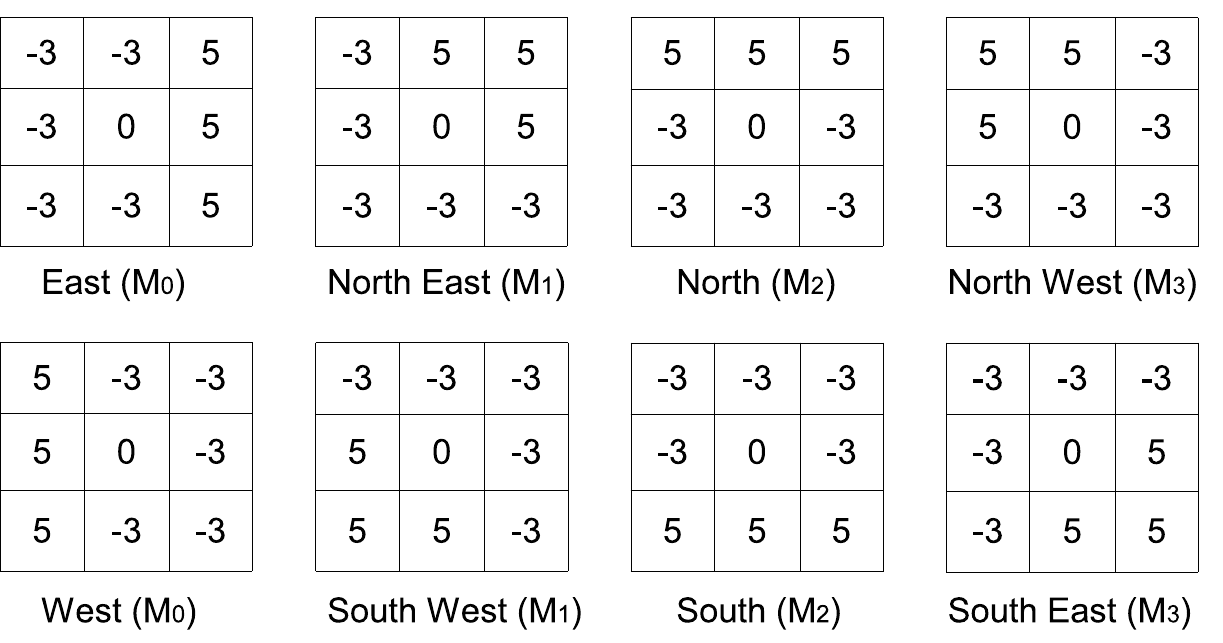}
\end{center}
\caption{Kirsch kernels utilized in the LDP approach.}
\label{kenels}
\end{figure}

\item \textit{Local ternary pattern (LTeP)}: Unlike LPH, LTeP does not encode the difference of power patterns in every kernel into 0 or 1, but encode them into other quantization values using a thresholding process \cite{Yuan2014}. Let consider $thr$ is the threshold parameter, $s_{c}$ presents the central power pattern in a patch of $3 \times 3$, and $s_{n}$ stands for the neighbor patterns, every central pattern $s_{c}^{\prime}$ can be encoded as follows: 

\begin{equation}
s_{c}^{\prime }=\left\{ 
\begin{array}{cc}
1 & \mathrm{if}~ s_{n} > s_{c} + thr \\ 
0 & \mathrm{if}~ s_{n} > s_{c} - thr ~ \mathrm{and}~s_{n} < s_{c} + thr \\ 
-1 & \mathrm{if}~ s_{n} < s_{c} - thr%
\end{array}%
\right. 
\end{equation}

\item \textit{Local Transitional Pattern (LTrP)}: It compares the transitions of intensity changes in small local regions (e.g. kernels of $3 \times 3$) in different orientations in order to binary encode the 2D representations of appliance power signals. Specifically, LTrP generates a bit (0/1) via the comparison the central power pattern of a $3 \times 3$ patch with only the intensities of two neighbors related to two particular directions \cite{Ahsan2013}.

\item \textit{Local binary pattern (LBP):} Is a texture descriptor that presents a low computation complexity along with a capability to capturing a good part of textural patterns of 2D representations. LBP represents micro-patterns in power matrices by an ensemble of simple computations around each power sample \cite{Tabatabaei2018}.

\item \textit{Binarized statistical image features (BSIF):} It constructs local descriptions of 2D representations via effectively encoding texture information and extracting histograms of local regions. Accordingly, binary codes for power patterns are extracted via the projection of local power regions onto a subspace, where basis-vectors were learnt using other natural images \cite{Kannala6460393}.

\end{itemize}

\begin{figure}[t!]
\begin{center}
\includegraphics[width=0.46\textwidth]{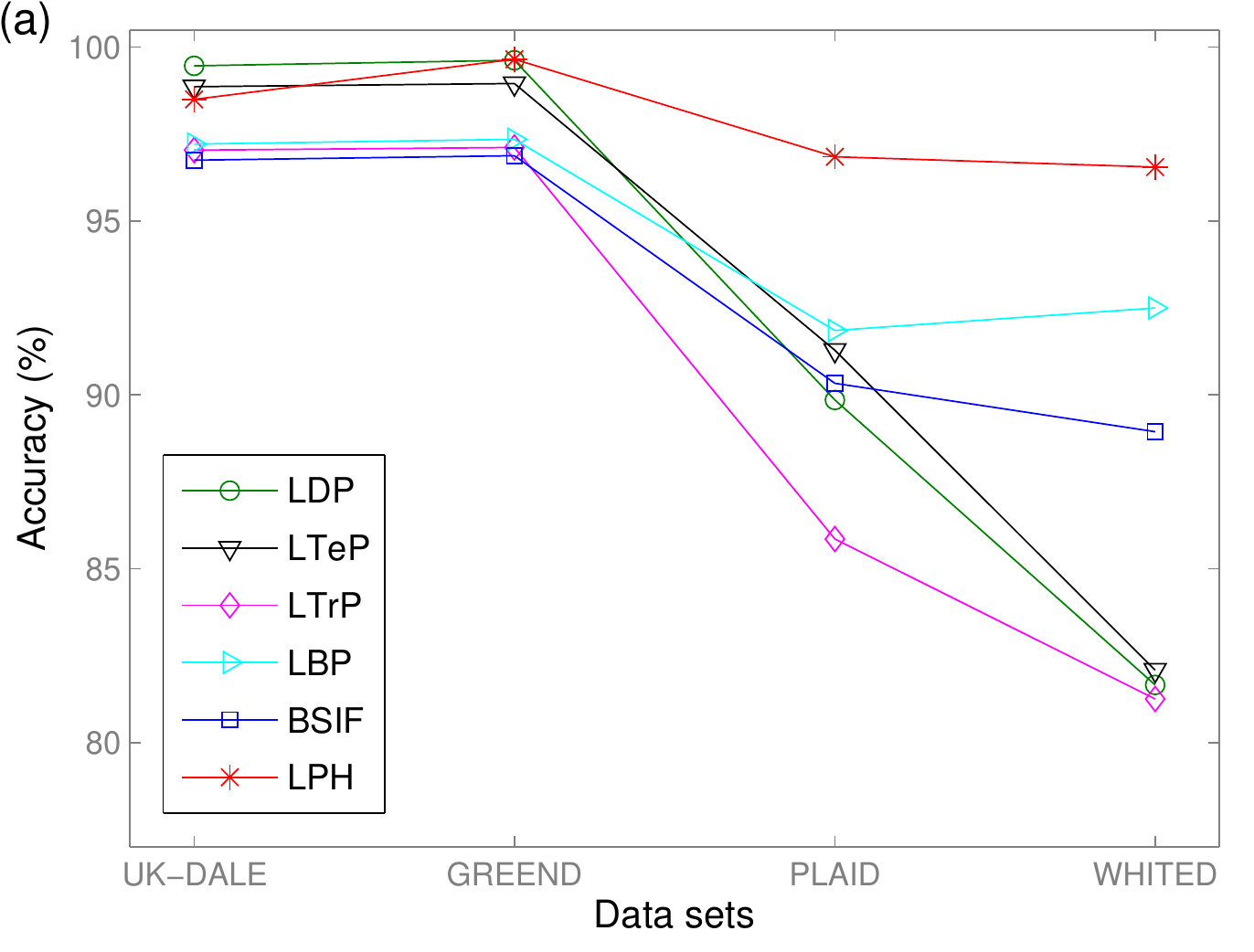}
\includegraphics[width=0.46\textwidth]{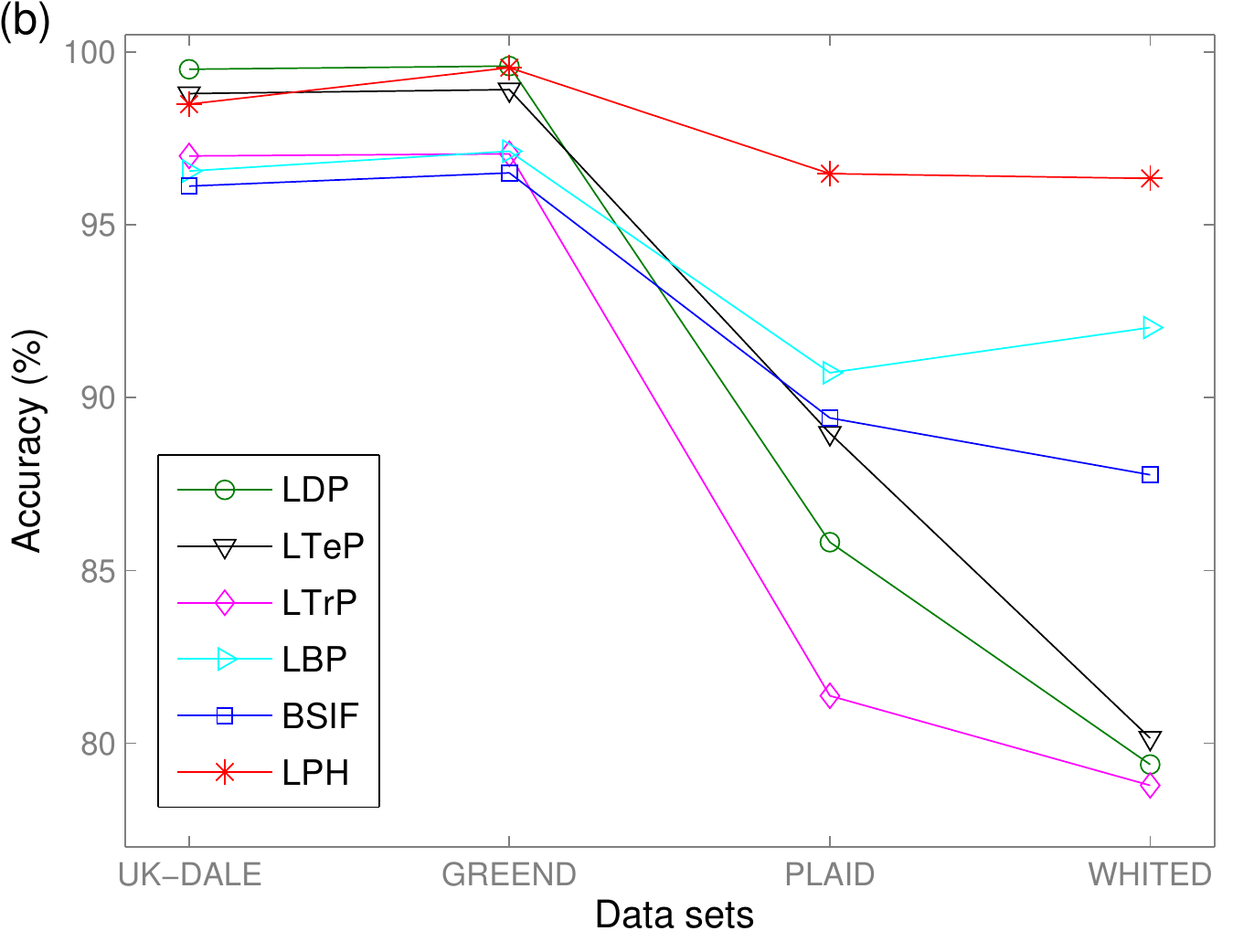}

\end{center}
\caption{The performance of LPH descriptor compared to other 2D feature extraction schemes in terms of (a) accuracy, and (b) F1 score.}
\label{LPHvariant}
\end{figure}

Table \ref{LPHvriants} along with Fig. \ref{LPHvariant} portray the performance of LPH in comparison with the aforementioned 2D descriptors, among them LBP, LDP, LTeP, BSIF and LTrP with regard to the histogram length, accuracy and F1 score. The results have been obtained by considering the IKNN for all descriptors (K=5). I has been evidently shown that high performance has been obtained with all the descriptors under UK-DALE. Explicitly, all the descriptors have achieved an accuracy and F1 score of more than 96\%. On the other hand, LDP and LTeP descriptors marginally surpass the LPH under this repository. On the contrary, the performance of the other descriptors have been highly dropped under PLAID and WHITED and only LPH maintains good accuracy and F1 score results. For instance, LPH has attained 96.85\% accuracy and 96.48 F1 score under PLAID and 96.55\% accuracy and	96.34\% F1 score under WHITED. In this context, under PLAID, LPH outperforms BSIF, LBP, LTrP, LTeP and LDP in terms of the accuracy by more than 6\%, 5\%, 11\%, 5.5\% and 7\%, respectively. While in terms of the F1 score, it outperforms them by 7\%, 5.5\%, 15\%, 7\%  and 10\%, respectively.

Conversely, the performance variation reported under the different data sets is due to frequency resolutions variation, in addition because UK-DALE records appliance power consumption for multiple days (i.e. it collects the consumption from the same devices but for distinct days for a long period)  while  PLAID and WHITED data sets observe different devices from distinct manufacturers (brands) and which are belonging to the same device category.   

\begin{table}[t!]
\caption{Performance of the LPH-based descriptor vs. other 2D descriptors with reference to the histogram length, accuracy and F1 score. }
\label{LPHvriants}
\begin{center}

\begin{tabular}{llllllllll}
\hline
{\small Descriptor} & {\small Histogram} & \multicolumn{2}{c}{\small UK-DALE}
& \multicolumn{2}{c}{GREEND} & \multicolumn{2}{c}{\small PLAID} & 
\multicolumn{2}{c}{\small WHITED} \\ \cline{3-10}
& {\small length} & {\small Accuracy} & {\small F1 score} & Accuracy
& F1 score & {\small Accuracy} & {\small F1 score} & {\small %
Accuracy} & {\small F1 score} \\ \hline
{\small LDP} & {\small 56} & {\small \textbf{99.46}} & {\small \textbf{99.50}} & 99.62
& \textbf{99.59} & {\small 89.79} & {\small 85.82} & {\small 81.66} & 
{\small 79.38} \\ 
{\small LTeP} & {\small 512} & {\small 98.86} & {\small 98.80} & 98.95 & 98.91 & {\small 91.28} & {\small 88.97} & {\small 82.08} & 
{\small 80.15} \\ 
{\small LTrP} & {\small 256} & {\small 97.04} & {\small 96.99} & 97.11 & 97.05 & {\small 85.81} & {\small 81.37} & {\small 81.25} & 
{\small 78.78} \\ 
{\small LBP} & {\small 256} & {\small 97.21} & {\small 96.56} & 97.35 & 97.13 & {\small 91.83} & {\small 90.72} & {\small 92.5} & 
{\small 92.03} \\ 
{\small BSIF} & {\small 256} & {\small 96.75} & {\small 96.12} & 96.88 & 96.50 & {\small 90.33} & {\small 89.41} & {\small 88.94} & 
{\small 87.77} \\ 
{\small LPH} & {\small 256} & {\small 98.51} & {\small 98.49} & \textbf{99.65} & 99.55 & {\small \textbf{96.85}} & {\small \textbf{96.48}} & {\small \textbf{96.55}} & 
{\small \textbf{96.34}} \\ \hline
\end{tabular}

\end{center}
\end{table}

Moving forward, we have evaluated the computation cost of the proposed appliance identification scheme based on different 2D descriptors in order to demonstrates its applicability in real-time scenarios. Accordingly, the computation time for the training and test phases of our approach have been computed using MATLAB 9.4. The computational costs are computed on a laptop having a Core i7-85500 with 32 GB RAM and 1.97 GHz. Table \ref{time} depicts the obtained computational times (in sec) with regard to various 2D descriptors under the three data sets adopted in this framework.

Accordingly, it has been clearly seen that the appliance identification based LPH achieves the lowest computational time in comparison with the
other descriptors for both the training and test stages under the three data sets. Moreover, the test time of LPH based solution under PLAID and WHITED is less than 1 sec, which can proves that it is possible to implement it for real-time applications since most of the transmitter can transmit data with a sampling rate of more than 1 sec. On the other flip, the test time of the LPH based solution has increased under UK-ALE to more than 2 sec because in this case long daily consumption signatures are analyzed. In contrast to PLAID and WHITED, where short appliance fingerprints from are considered.

\begin{table}[t!]
\caption{Computational time (in sec) of the proposed appliance identification based on different 2D descriptors. }
\label{time}
\begin{center}

\begin{tabular}{lcccccccc}
\hline
& \multicolumn{8}{c}{\small Time (in sec)} \\ \cline{2-9}
{\small 2D descriptors \ \ } & \multicolumn{2}{c}{\small UK-DALE} & 
\multicolumn{2}{c}{GREEND} & \multicolumn{2}{c}{\small PLAID} & 
\multicolumn{2}{c}{\small WHITED} \\ \cline{2-9}
& \ {\small training \ } & \ \ \ \ {\small test \ \ \ \ } &  training & \ \ \ \ test \ \ \  & \ {\small training \ } & \ \ \ \ \ 
{\small test \ \ \ } & \ {\small training \ } & \ \ \ \ {\small test \ \ \ \ 
} \\ \hline
{\small LDP} & {\small 25.18} & {\small 3.71} & 32.23 & 4.73 & {\small 8.89} & {\small 1.27} & {\small 6.17} & {\small 0.88} \\ 
{\small LTeP} & {\small 31.22} & {\small 3.86} & 39.97 & 4.96 & {\small 11.39} & {\small 1.69} & {\small 7.76} & {\small 1.19} \\ 
{\small LTrP} & {\small 34.69} & {\small 4.38} & 44.41 & 5.61 & {\small 12.48} & {\small 1.44} & {\small 8.42} & {\small 1.03} \\ 
{\small LBP} & {\small 19.55} & {\small 2.96} & 25.44 & 3.77 & {\small 6.26} & {\small 0.97} & {\small 4.13} & {\small 0.69} \\ 
{\small BSIF} & {\small 39.17} & {\small 5.11} & 49.17 & 6.61 & {\small 13.75} & {\small 1.76} & {\small 9.27} & {\small 1.25} \\ 

{\small LPH} & {\small \textbf{19.45}} & {\small \textbf{2.92}} & \textbf{21.89} & \textbf{3.76} & {\small \textbf{5.91}} & {\small \textbf{0.93}} & {\small \textbf{3.68}} & {\small \textbf{0.64}} \\ \hline
\end{tabular}

\end{center}
\end{table}

\subsection{Comparison with existing load identification frameworks} 
Table \ref{AIScomp} recapitulates the results of various existing load identification frameworks under REDD data set, in comparison with the proposed LPH-IKNN solution and with reference to different parameters, among them the description of learning model, its type, number of the device categories and accuracy performance. It has been clearly seen that the LPH-IKNN framework outperforms all other architectures considered in this study. Moreover, LPH-IKNN has a low computational cost, which can make it a candidate for real-time applications. On the other side, it is worth noting that the proposed method is evaluated under three distinct power repositories with different sampling rates, in which it presents promising results. In contrast, each of the other methods is only validated under one data set, therefore, this increases the credibility of the our study and proves that it could be deployed under different scenarios without caring about the sampling rate.

\begin{table}[t!]
\caption{Performance of the proposed LPH-IKNN based load identification system vs. existing solutions with reference to different criteria.}
\label{AIScomp}
\begin{center}

\begin{tabular}{ccccc}
\hline
{\small Framework} & {\small Approach} & {\small Learning } & {\small \ \ \#
appliance \ \ \ } & {\small \ \ \ Accuracy \ \ \ } \\ 
&  & {\small type} & {\small classes} & {\small (\%)} \\ \hline
\multicolumn{1}{l}{{\small \cite{HIMEUR2020114877}}} & {\small MSWPT + DBT}
& {\small supervised} & {\small 9} & {\small 96.81} \\ 
\multicolumn{1}{l}{{\small \cite{Park2019}}} & {\small ANN} & {\small %
supervised} & {\small 8} & {\small 83.8} \\ 
\multicolumn{1}{l}{{\small \cite{Ma8118142}}} & {\small fingerprint-weighting KNN%
} & {\small supervised } & {\small 6} & {\small 83.25} \\ 
\multicolumn{1}{l}{{\small \cite{Guedes2015}}} & {\small HOS} & {\small %
supervised} & {\small 11} & {\small 96.8} \\ 
\multicolumn{1}{l}{{\small \cite{Wang2012}}} & {\small \ \ \ mean-shift
clustering \ \ \ \ } & {\small \ \ unsupervied \ \ } & {\small 13} & {\small %
80} \\ 
\multicolumn{1}{l}{{\small \cite{Dinesh2016}}} & {\small Karhunen Lo\'{e}ve}
& {\small supervised} & {\small N/A} & {\small 87} \\ 
\multicolumn{1}{l}{{\small \cite{Morais2019}}} & {\small AANN } & {\small %
supervised} & {\small 5} & {\small 97.7} \\ 
\multicolumn{1}{l}{{\small \cite{Zhiren2019}}} & {\small AdaBoost} & {\small %
supervised} & {\small 5} & {\small 94.8} \\ 
\multicolumn{1}{l}{{\small \cite{Ghosh2019}}} & {\small Fuzzy model} & 
{\small supervised} & {\small 7} & {\small 91.5} \\ 
\multicolumn{1}{l}{{\small \cite{YAN2019101393} \ \ }} & {\small Bayesian
classifier + correlation} & {\small supervised} & {\small 29} & {\small 95.6} \\ 
%& {\small + correlation} &  &  &  \\ 
\multicolumn{1}{l}{\small Our} & {\small LPH + IKNN} & {\small supervised} & 
{\small 9} & {\small 98.5} \\ \hline
\end{tabular}

\end{center}
\end{table}

All in all, it is of significant importance to mention that the proposed LPH-IKNN has been validated using four different data sets (i.e. UK-DALE, GREEND, PLAID and UK-DALE) including different kinds of power signatures recorded (i) with distinct frequency resolutions, and (ii) under different scenarios. For instance, under both UK-DALE and GREEND, we collect the power consumption signatures that vary through the time for a set of appliances (i.e. each daily consumption trace represents a power signature); while under both PLAID and WHITED, for each appliance category, the power traces are gleaned from different manufacturers. In this regard, validating our solution under these data sets using different scenarios, has helped in (i) showing its high performance although the frequency resolution has been changed, and (ii) proving its capability to be implemented in real-application scenarios since it can identify appliance-level data even if they are from different manufacturers and although the power signatures change from a day to another.

\section{Conclusion} \label{sec5}

In this paper, a novel method for performing accurate appliance identification and hence improving the performance of the NILM systems has been presented. The applicability of a local 2D descriptor, namely LPH-IKNN, to recognize electrical devices has been successfully validated. Consequently, other types of 2D descriptors can be investigated in order to further improve the identification accuracy, such as local texture descriptors, color histograms, moment-based descriptors and scale-invariant descriptors. This line of research is full of challenges and plenty of opportunities are available. Moving forward, in addition to the high performance reached, the LPH-IKNN based appliance identifications scheme has shown a low computational cost because of the use of a fast 2D descriptor along with the IKNN, which uses a smart strategy to reduce the training and test times. Furthermore, LPH-IKNN acts also as a dimensionality reduction, in which very short histograms have been derived to represent appliance fingerprints.

On the other hand, although LPH-IKNN has shown very promising performance, it still has some drawbacks among them is its reliance on a supervised learning process. Explicitly, this could limit its application in some scenarios, where it might be difficult to collect data to train the proposed model. To that end, it is part of our next future work to change the learning process by building an improved version of this LPH-IKNN using an unsupervised learning approach. Moreover, another option is by adding a transfer learning module to eliminate the need to collect new data for training our system if the sampling frequency of collected data is changed. Moreover, IKNN classifier could be replaced by any other improved algorithm that enables an automatic selection of the $k$ value to simplify the use of LPH-IKNN in real application scenarios. In this context, the GBKNN classifier discussed in Section \ref{sec221} seems to be a good alternative that could be investigated in our future work.

On the other hand, due to the size of appliance identification based data sets is not very large, it will be of significant importance to investigate the use of other feature extraction methods in our future work, which are very convenient for small data sets, e.g. rough set based techniques \cite{xia2020gbnrs,xia2020lra}. The latter helps also in attribute reduction and feature selection and hence it could further reduce the computational cost of the appliance identification task to support real-time applications. Finally, it will also be part of our future work to focus on developing a powerful recommender system, which can use the output of the LPH-IKNN based NILM system to detect abnormal power consumption in buildings before triggering tailored recommendations to help end-users in reducing wasted energy.

\section*{Acknowledgements}\label{acknowledgements}
This paper was made possible by National Priorities Research Program (NPRP) grant No. 10-0130-170288 from the Qatar National Research Fund (a member of Qatar Foundation). The statements made herein are solely the responsibility of the authors.

%\section*{References}


\begin{thebibliography}{10}

\bibitem{Elattar2020}
E.~Elattar, N.~Sabiha, M.~Alsharef, M.~Metwaly, A.~Abd-Elhady, I.~Taha, Short
  term electric load forecasting using hybrid algorithm for smart cities,
  Applied Intelligence (2020) 1--21\href
  {http://dx.doi.org/10.1007/s10489-020-01728-x}
  {\path{doi:10.1007/s10489-020-01728-x}}.

\bibitem{sardianos2020emergence}
C.~Sardianos, I.~Varlamis, C.~Chronis, G.~Dimitrakopoulos, A.~Alsalemi,
  Y.~Himeur, F.~Bensaali, A.~Amira, The emergence of explainability of
  intelligent systems: Delivering explainable and personalized recommendations
  for energy efficiency, International Journal of Intelligent Systems 36 (2021)
  656--680.

\bibitem{REHABC2020}
C.~Sardianos, I.~Varlamis, G.~Dimitrakopoulos, D.~Anagnostopoulos, A.~Alsalemi,
  F.~Bensaali, Y.~Himeur, A.~Amira, Rehab-c: Recommendations for energy habits
  change, Future Generation Computer Systems 112 (2020) 394 -- 407.

\bibitem{alsalemi2020micro}
A.~Alsalemi, Y.~Himeur, F.~Bensaali, A.~Amira, C.~Sardianos, C.~Chronis,
  I.~Varlamis, G.~Dimitrakopoulos, A micro-moment system for domestic energy
  efficiency analysis, IEEE Systems Journal.

\bibitem{HIMEUR2020114877}
Y.~Himeur, A.~Alsalemi, F.~Bensaali, A.~Amira, Robust event-based non-intrusive
  appliance recognition using multi-scale wavelet packet tree and ensemble
  bagging tree, Applied Energy 267 (2020) 114877.

\bibitem{PEREIRA2020102399}
L.~Pereira, N.~Nunes, An empirical exploration of performance metrics for event
  detection algorithms in non-intrusive load monitoring, Sustainable Cities and
  Society 62 (2020) 102399.

\bibitem{LIU2020101918}
H.~Liu, C.~Yu, H.~Wu, C.~Chen, Z.~Wang, An improved non-intrusive load
  disaggregation algorithm and its application, Sustainable Cities and Society
  53 (2020) 101918.

\bibitem{alsalemi2020achieving}
A.~Alsalemi, Y.~Himeur, F.~Bensaali, A.~Amira, C.~Sardianos, I.~Varlamis,
  G.~Dimitrakopoulos, Achieving domestic energy efficiency using micro-moments
  and intelligent recommendations, IEEE Access 8 (2020) 15047--15055.

\bibitem{Welikala8039522}
S.~{Welikala}, C.~{Dinesh}, M.~P.~B. {Ekanayake}, R.~I. {Godaliyadda},
  J.~{Ekanayake}, Incorporating appliance usage patterns for non-intrusive load
  monitoring and load forecasting, IEEE Transactions on Smart Grid 10~(1)
  (2019) 448--461.

\bibitem{He8669739}
K.~{He}, D.~{Jakovetic}, B.~{Zhao}, V.~{Stankovic}, L.~{Stankovic}, S.~{Cheng},
  A generic optimisation-based approach for improving non-intrusive load
  monitoring, IEEE Transactions on Smart Grid 10~(6) (2019) 6472--6480.

\bibitem{Park2019}
S.~W. {Park}, L.~B. {Baker}, P.~D. {Franzon}, Appliance identification
  algorithm for a non-intrusive home energy monitor using cogent confabulation,
  IEEE Transactions on Smart Grid 10~(1) (2019) 714--721.

\bibitem{Ma8118142}
M.~{Ma}, W.~{Lin}, J.~{Zhang}, P.~{Wang}, Y.~{Zhou}, X.~{Liang}, Toward
  energy-awareness smart building: Discover the fingerprint of your electrical
  appliances, IEEE Transactions on Industrial Informatics 14~(4) (2018)
  1458--1468.

\bibitem{Makonin7317784}
S.~{Makonin}, F.~{Popowich}, I.~V. {Bajić}, B.~{Gill}, L.~{Bartram},
  Exploiting hmm sparsity to perform online real-time nonintrusive load
  monitoring, IEEE Transactions on Smart Grid 7~(6) (2016) 2575--2585.

\bibitem{Guedes2015}
J.~D.~S. Guedes, D.~D. Ferreira, B.~H.~G. Barbosa, C.~A. Duque, A.~S.
  Cerqueira, Non-intrusive appliance load identification based on higher-order
  statistics, IEEE Latin America Transactions 13 (2015) 3343--3349.

\bibitem{Ji8684887}
T.~Y. {Ji}, L.~{Liu}, T.~S. {Wang}, W.~B. {Lin}, M.~S. {Li}, Q.~H. {Wu},
  Non-intrusive load monitoring using additive factorial approximate maximum a
  posteriori based on iterative fuzzy {c-Means}, IEEE Transactions on Smart
  Grid 10~(6) (2019) 6667--6677.

\bibitem{He7539273}
K.~{He}, L.~{Stankovic}, J.~{Liao}, V.~{Stankovic}, Non-intrusive load
  disaggregation using graph signal processing, IEEE Transactions on Smart Grid
  9~(3) (2018) 1739--1747.

\bibitem{Li8437176}
D.~{Li}, S.~{Dick}, Residential household non-intrusive load monitoring via
  graph-based multi-label semi-supervised learning, IEEE Transactions on Smart
  Grid 10~(4) (2019) 4615--4627.

\bibitem{Zhao2018Access}
B.~Zhao, K.~He, L.~Stankovic, V.~Stankovic, Improving event-based non-intrusive
  load monitoring using graph signal processing, IEEE Access PP (2018) 1--1.

\bibitem{Kolter2010EDV}
J.~Z. Kolter, S.~Batra, A.~Y. Ng, Energy disaggregation via discriminative
  sparse coding, in: Proceedings of the 23rd International Conference on Neural
  Information Processing Systems - Volume 1, NIPS'10, Curran Associates Inc.,
  USA, 2010, pp. 1153--1161.

\bibitem{Singh2019SG}
S.~{Singh}, A.~{Majumdar}, Analysis co-sparse coding for energy disaggregation,
  IEEE Transactions on Smart Grid 10~(1) (2019) 462--470.

\bibitem{Singh7847445}
S.~{Singh}, A.~{Majumdar}, Deep sparse coding for non-intrusive load
  monitoring, IEEE Transactions on Smart Grid 9~(5) (2018) 4669--4678.

\bibitem{Rahimpour7835299}
A.~{Rahimpour}, H.~{Qi}, D.~{Fugate}, T.~{Kuruganti}, Non-intrusive energy
  disaggregation using non-negative matrix factorization with sum-to-k
  constraint, IEEE Transactions on Power Systems 32~(6) (2017) 4430--4441.

\bibitem{HAMMAD2019180}
M.~Hammad, S.~Zhang, K.~Wang, A novel two-dimensional ecg feature extraction
  and classification algorithm based on convolution neural network for human
  authentication, Future Generation Computer Systems 101 (2019) 180 -- 196.

\bibitem{Du7130652}
L.~{Du}, D.~{He}, R.~G. {Harley}, T.~G. {Habetler}, Electric load
  classification by binary voltage-current trajectory mapping, IEEE
  Transactions on Smart Grid 7~(1) (2016) 358--365.

\bibitem{Gao7418189}
J.~{Gao}, E.~C. {Kara}, S.~{Giri}, M.~{Bergés}, A feasibility study of
  automated plug-load identification from high-frequency measurements, in: 2015
  IEEE Global Conference on Signal and Information Processing (GlobalSIP),
  2015, pp. 220--224.

\bibitem{Liu8580416}
Y.~{Liu}, X.~{Wang}, W.~{You}, Non-intrusive load monitoring by voltage-current
  trajectory enabled transfer learning, IEEE Transactions on Smart Grid 10~(5)
  (2019) 5609--5619.

\bibitem{DEBAETS2019645}
L.~D. Baets, C.~Develder, T.~Dhaene, D.~Deschrijver, Detection of unidentified
  appliances in non-intrusive load monitoring using siamese neural networks,
  International Journal of Electrical Power \& Energy Systems 104 (2019) 645 --
  653.

\bibitem{JUNKER2018175}
R.~G. Junker, A.~G. Azar, R.~A. Lopes, K.~B. Lindberg, G.~Reynders, R.~Relan,
  H.~Madsen, Characterizing the energy flexibility of buildings and districts,
  Applied Energy 225 (2018) 175--182.

\bibitem{Himeur2020icict}
Y.~Himeur, A.~Alsalemi, F.~Bensaali, A.~Amira, Improving in-home appliance
  identification using fuzzy-neighbors-preserving analysis based
  qr-decomposition, in: International Congress on Information and Communication
  Technology, Springer, 2020, pp. 303--311.

\bibitem{WANG2020114145}
X.~Wang, S.-H. Ahn, Real-time prediction and anomaly detection of electrical
  load in a residential community, Applied Energy 259 (2020) 114145.

\bibitem{himeur2020effective}
Y.~Himeur, A.~Alsalemi, F.~Bensaali, A.~Amira, Effective non-intrusive load
  monitoring of buildings based on a novel multi-descriptor fusion with
  dimensionality reduction, Applied Energy 279 (2020) 115872.

\bibitem{Himeur2020iscas}
Y.~Himeur, A.~Alsalemi, F.~Bensaali, A.~Amira, Efficient multi-descriptor
  fusion for non-intrusive appliance recognition, in: 2020 IEEE International
  Symposium on Circuits and Systems (ISCAS), IEEE, 2020, pp. 1--5.

\bibitem{shi2019nonintrusive}
X.~Shi, H.~Ming, S.~Shakkottai, L.~Xie, J.~Yao, Nonintrusive load monitoring in
  residential households with low-resolution data, Applied Energy 252 (2019)
  113283.

\bibitem{Himeur2020IJIS-AD}
Y.~{Himeur}, A.~{Elsalemi}, F.~{Bensaali}, A.~Amira, Smart power consumption
  abnormality detection in buildings using micro-moments and improved k-nearest
  neighbors, Intenational Journal of Intelligent Systems (2020) 1--25.

\bibitem{mulongo2020anomaly}
J.~Mulongo, M.~Atemkeng, T.~Ansah-Narh, R.~Rockefeller, G.~M. Nguegnang, M.~A.
  Garuti, Anomaly detection in power generation plants using machine learning
  and neural networks, Applied Artificial Intelligence 34~(1) (2020) 64--79.

\bibitem{himeur2020novel}
Y.~Himeur, A.~Alsalemi, F.~Bensaali, A.~Amira, A novel approach for detecting
  anomalous energy consumption based on micro-moments and deep neural networks,
  Cognitive Computation 12~(6) (2020) 1381--1401.

\bibitem{mehta2018new}
S.~Mehta, X.~Shen, J.~Gou, D.~Niu, A new nearest centroid neighbor classifier
  based on k local means using harmonic mean distance, Information 9~(9) (2018)
  234.

\bibitem{abu2019effects}
H.~A. Abu~Alfeilat, A.~B. Hassanat, O.~Lasassmeh, A.~S. Tarawneh, M.~B.
  Alhasanat, H.~S. Eyal~Salman, V.~S. Prasath, Effects of distance measure
  choice on k-nearest neighbor classifier performance: A review, Big data 7~(4)
  (2019) 221--248.

\bibitem{gou2019generalized}
J.~Gou, H.~Ma, W.~Ou, S.~Zeng, Y.~Rao, H.~Yang, A generalized mean
  distance-based k-nearest neighbor classifier, Expert Systems with
  Applications 115 (2019) 356--372.

\bibitem{gou2019local}
J.~Gou, W.~Qiu, Z.~Yi, Y.~Xu, Q.~Mao, Y.~Zhan, A local mean
  representation-based k-nearest neighbor classifier, ACM Transactions on
  Intelligent Systems and Technology (TIST) 10~(3) (2019) 1--25.

\bibitem{gou2019locality}
J.~Gou, W.~Qiu, Z.~Yi, X.~Shen, Y.~Zhan, W.~Ou, Locality constrained
  representation-based k-nearest neighbor classification, Knowledge-Based
  Systems 167 (2019) 38--52.

\bibitem{xia2019granular}
S.~Xia, Y.~Liu, X.~Ding, G.~Wang, H.~Yu, Y.~Luo, Granular ball computing
  classifiers for efficient, scalable and robust learning, Information Sciences
  483 (2019) 136--152.

\bibitem{chui2013appliance}
K.~T. Chui, K.~Tsang, S.~Chung, L.~Yeung, Appliance signature identification
  solution using k-means clustering, in: IECON 2013-39th Annual Conference of
  the IEEE Industrial Electronics Society, IEEE, 2013, pp. 8420--8425.

\bibitem{henriques2020combining}
J.~Henriques, F.~Caldeira, T.~Cruz, P.~Sim{\~o}es, Combining k-means and
  xgboost models for anomaly detection using log datasets, Electronics 9~(7)
  (2020) 1164.

\bibitem{saba2020recent}
T.~Saba, Recent advancement in cancer detection using machine learning:
  Systematic survey of decades, comparisons and challenges, Journal of
  Infection and Public Health 13~(9) (2020) 1274--1289.

\bibitem{alsayat2016social}
A.~Alsayat, H.~El-Sayed, Social media analysis using optimized k-means
  clustering, in: 2016 IEEE 14th International Conference on Software
  Engineering Research, Management and Applications (SERA), IEEE, 2016, pp.
  61--66.

\bibitem{yu2018two}
S.-S. Yu, S.-W. Chu, C.-M. Wang, Y.-K. Chan, T.-C. Chang, Two improved k-means
  algorithms, Applied Soft Computing 68 (2018) 747--755.

\bibitem{zhang2018improved}
G.~Zhang, C.~Zhang, H.~Zhang, Improved k-means algorithm based on density
  canopy, Knowledge-based systems 145 (2018) 289--297.

\bibitem{lu2019improved}
W.~Lu, Improved k-means clustering algorithm for big data mining under hadoop
  parallel framework, Journal of Grid Computing (2019) 1--12.

\bibitem{whang2015non}
J.~J. Whang, I.~S. Dhillon, D.~F. Gleich, Non-exhaustive, overlapping k-means,
  in: Proceedings of the 2015 SIAM International Conference on Data Mining,
  SIAM, 2015, pp. 936--944.

\bibitem{khanmohammadi2017improved}
S.~Khanmohammadi, N.~Adibeig, S.~Shanehbandy, An improved overlapping k-means
  clustering method for medical applications, Expert Systems with Applications
  67 (2017) 12--18.

\bibitem{Tao2019}
H.~Tao, Detecting smoky vehicles from traffic surveillance videos based on
  dynamic features, Applied Intelligence 50 (2019) 1--16.
\newblock \href {http://dx.doi.org/10.1007/s10489-019-01589-z}
  {\path{doi:10.1007/s10489-019-01589-z}}.

\bibitem{Kumar9097394}
A.~{Kumar}, S.~K. {Singh}, S.~{Saxena}, A.~K. {Singh}, S.~{Shrivastava},
  K.~{Lakshmanan}, N.~{Kumar}, R.~K. {Singh}, Comhisp: A novel feature
  extractor for histopathological image classification based on fuzzy svm with
  within-class relative density, IEEE Transactions on Fuzzy Systems (2020)
  1--1.

\bibitem{Gong7812744}
D.~{Gong}, Z.~{Li}, W.~{Huang}, X.~{Li}, D.~{Tao}, Heterogeneous face
  recognition: A common encoding feature discriminant approach, IEEE
  Transactions on Image Processing 26~(5) (2017) 2079--2089.

\bibitem{Ramirez8681394}
D.~{Valdes-Ramirez}, M.~A. {Medina-Pérez}, R.~{Monroy}, O.~{Loyola-González},
  J.~{Rodríguez}, A.~{Morales}, F.~{Herrera}, A review of fingerprint feature
  representations and their applications for latent fingerprint identification:
  Trends and evaluation, IEEE Access 7 (2019) 48484--48499.

\bibitem{Batra:2019:TRS}
N.~Batra, R.~Kukunuri, A.~Pandey, R.~Malakar, R.~Kumar, O.~Krystalakos,
  M.~Zhong, P.~Meira, O.~Parson, Towards reproducible state-of-the-art energy
  disaggregation, in: Proceedings of the 6th ACM International Conference on
  Systems for Energy-Efficient Buildings, Cities, and Transportation, BuildSys
  '19, ACM, New York, NY, USA, 2019, pp. 193--202.

\bibitem{Batra:2019:DRS}
N.~Batra, R.~Kukunuri, A.~Pandey, R.~Malakar, R.~Kumar, O.~Krystalakos,
  M.~Zhong, P.~Meira, O.~Parson, A demonstration of reproducible
  state-of-the-art energy disaggregation using nilmtk, in: Proceedings of the
  6th ACM International Conference on Systems for Energy-Efficient Buildings,
  Cities, and Transportation, BuildSys '19, ACM, New York, NY, USA, 2019, pp.
  358--359.

\bibitem{Lu8090442}
T.~{Lu}, Z.~{Xu}, B.~{Huang}, An event-based nonintrusive load monitoring
  approach: Using the simplified viterbi algorithm, IEEE Pervasive Computing
  16~(4) (2017) 54--61.

\bibitem{Batra2019}
N.~Batra, R.~Kukunuri, A.~Pandey, R.~Malakar, R.~Kumar, O.~Krystalakos,
  M.~Zhong, P.~Meira, O.~Parson, A demonstration of reproducible
  state-of-the-art energy disaggregation using nilmtk, in: Proceedings of the
  6th ACM International Conference on Systems for Energy-Efficient Buildings,
  Cities, and Transportation, BuildSys ’19, Association for Computing
  Machinery, New York, NY, USA, 2019, pp. 358--359.

\bibitem{Batra2014ACM}
N.~Batra, J.~Kelly, O.~Parson, H.~Dutta, W.~Knottenbelt, A.~Rogers, A.~Singh,
  M.~Srivastava, Nilmtk: An open source toolkit for non-intrusive load
  monitoring, 2014.

\bibitem{Liu2011}
W.~Liu, S.~Chawla, Class confidence weighted knn algorithms for imbalanced data
  sets, Vol. 6635, 2011, pp. 345--356.
\newblock \href {http://dx.doi.org/10.1007/978-3-642-20847-8_29}
  {\path{doi:10.1007/978-3-642-20847-8_29}}.

\bibitem{Zhang7858565}
B.~{Zang}, R.~{Huang}, L.~{Wang}, J.~{Chen}, F.~{Tian}, X.~{Wei}, An improved
  knn algorithm based on minority class distribution for imbalanced dataset,
  in: 2016 International Computer Symposium (ICS), 2016, pp. 696--700.

\bibitem{UK-DALE2015}
J.~Kelly, W.~Knottenbelti, The uk-dale dataset, domestic appliance-level
  electricity demand and whole-house demand from five uk homes, Scientific Data
  2~(150007) (2015) 1 -- 14.

\bibitem{GREEND2014}
A.~{Monacchi}, D.~{Egarter}, W.~{Elmenreich}, S.~{D'Alessandro}, A.~M.
  {Tonello}, Greend: An energy consumption dataset of households in italy and
  austria, in: 2014 IEEE International Conference on Smart Grid Communications
  (SmartGridComm), 2014, pp. 511--516.

\bibitem{PLAID2014}
J.~Gao, S.~Giri, E.~C. Kara, M.~Berg{\'e}s, Plaid: A public dataset of
  high-resolution electrical appliance measurements for load identification
  research, in: Proceedings of the 1st ACM Conference on Embedded Systems for
  Energy-Efficient Buildings, BuildSys '14, ACM, New York, NY, USA, 2014, pp.
  198--199.

\bibitem{WHITED2016}
M.~Kahl, A.~U. Haq, T.~Kriechbaumer, H.-A. Jacobsen, Whited-a worldwide
  household and industry transient energy data set, in: 3rd International
  Workshop on Non-Intrusive Load Monitoring, 2016.

\bibitem{Perumal2016}
P.~R. Srinivasa, P.~C. Mouli, Dimensionality reduced local directional pattern
  {(DR-LDP)} for face recognition, Expert Systems with Applications 63 (2016)
  66--73.

\bibitem{Yuan2014}
J.-H. Yuan, H.-D. Zhu, Y.~Gan, L.~Shang, Enhanced local ternary pattern for
  texture classification, in: D.-S. Huang, V.~Bevilacqua, P.~Premaratne (Eds.),
  Intelligent Computing Theory, Springer International Publishing, Cham, 2014,
  pp. 443--448.

\bibitem{Ahsan2013}
T.~Ahsan, T.~Jabid, U.-P. Chong, Facial expression recognition using local
  transitional pattern on gabor filtered facial images, IETE Technical Review
  30~(1) (2013) 47--52.

\bibitem{Tabatabaei2018}
S.~M. {Tabatabaei}, A.~{Chalechale}, One dimensional second order derivative
  local binary pattern for hand gestures classification using semg signals, in:
  2018 8th International Conference on Computer and Knowledge Engineering
  (ICCKE), 2018, pp. 16--19.

\bibitem{Kannala6460393}
J.~{Kannala}, E.~{Rahtu}, Bsif: Binarized statistical image features, in:
  Proceedings of the 21st International Conference on Pattern Recognition
  (ICPR2012), 2012, pp. 1363--1366.

\bibitem{Wang2012}
Z.~{Wang}, G.~{Zheng}, Residential appliances identification and monitoring by
  a nonintrusive method, IEEE Transactions on Smart Grid 3~(1) (2012) 80--92.

\bibitem{Dinesh2016}
C.~{Dinesh}, B.~W. {Nettasinghe}, R.~I. {Godaliyadda}, M.~P.~B. {Ekanayake},
  J.~{Ekanayake}, J.~V. {Wijayakulasooriya}, Residential appliance
  identification based on spectral information of low frequency smart meter
  measurements, IEEE Transactions on Smart Grid 7~(6) (2016) 2781--2792.

\bibitem{Morais2019}
L.~R. {Morais}, A.~R.~G. {Castro}, Competitive autoassociative neural networks
  for electrical appliance identification for non-intrusive load monitoring,
  IEEE Access 7 (2019) 111746--111755.

\bibitem{Zhiren2019}
R.~{Zhiren}, T.~{Bo}, W.~{Longfeng}, L.~{Hui}, L.~{Yanfei}, W.~{Haiping},
  Non-intrusive load identification method based on integrated intelligence
  strategy, in: 2019 25th International Conference on Automation and Computing
  (ICAC), 2019, pp. 1--6.

\bibitem{Ghosh2019}
S.~{Ghosh}, A.~{Chatterjee}, D.~{Chatterjee}, Improved non-intrusive
  identification technique of electrical appliances for a smart residential
  system, IET Generation, Transmission Distribution 13~(5) (2019) 695--702.

\bibitem{YAN2019101393}
D.~Yan, Y.~Jin, H.~Sun, B.~Dong, Z.~Ye, Z.~Li, Y.~Yuan, Household appliance
  recognition through a bayes classification model, Sustainable Cities and
  Society 46 (2019) 101393.

\bibitem{xia2020gbnrs}
S.~Xia, Z.~Zhang, W.~Li, G.~Wang, E.~Giem, Z.~Chen, Gbnrs: A novel rough set
  algorithm for fast adaptive attribute reduction in classification, IEEE
  Transactions on Knowledge and Data Engineering.

\bibitem{xia2020lra}
S.~Xia, W.~Li, G.~Wang, X.~Gao, C.~Zhang, E.~Giem, Lra: an accelerated rough
  set framework based on local redundancy of attribute for feature selection,
  arXiv preprint arXiv:2011.00215.

\end{thebibliography}
\end{document}